\documentclass{ptephy}

\usepackage{graphicx}

\begin{document}

\title{Mode-coupling theory and beyond: a diagrammatic approach}

\author{\name{Grzegorz Szamel}{1}}
\address{\affil{1}
{Department of Chemistry, Colorado State University, Fort Collins, CO 80523, USA}
\email{grzegorz.szamel@colostate.edu}}

\begin{abstract}
For almost thirty years, mode-coupling theory has been the most widely discussed
and used but also the most controversial theory of the glass transition.
In this paper we briefly review the reasons for both its popularity and 
its controversy. We emphasize the need for the development of approaches that 
would be able to evaluate corrections to and extensions of the existing (standard) 
mode-coupling theory. Next, we review our diagrammatic formulation of the dynamics of 
interacting Brownian particles. We show that within this approach the standard
mode-coupling theory can be derived in a very simple way. Finally, we use our
diagrammatic approach to calculate two corrections to the 
mode-coupling theory's expression for the so-called irreducible memory function.
These corrections involve re-summations of well defined classes of non-mode-coupling
diagrams.
\end{abstract}

\subjectindex{xxxx, xxx} 

\maketitle


\section{Introduction}\label{sec:intro}

Since the publication, almost thirty years ago, of three nearly coincidental 
papers by Leutheusser \cite{Leutheusser}, Begtzelius, G\"otze and Sj\"olander \cite{BGS},
and Das, Mazenko, Ramaswamy and Toner \cite{DMRT}, mode-coupling theory has been 
the most widely used and discussed, but also the most controversial theoretical 
approach to the glass transition problem.  
One reason for the popularity of this theory was that 
during most of the last thirty years it 
was the only fully microscopic theory of glassy dynamics. 
To be more precise, it was the only theory that, 
at least for particles interacting via a spherically symmetric pair-wise additive 
potential, allowed one to start from the microscopic description of a glassy system 
(\textit{i.e.} the inter-particle interactions encoded 
in the pair correlation function or the static structure factor) and 
make predictions for dynamic quantities that can be measured in computer simulations 
or in real experiments. 
Importantly, to make these predictions the theory did not need, nor did it allow, 
using any fitting parameters. Thus, 
the mode-coupling theory was easily testable and falsifiable. For this reason, it 
stimulated a great number of simulational \cite{KobLH} 
and experimental \cite{GoetzeJPCM1999} studies that intended to verify its predictions. 
Furthermore, simplified versions of the mode-coupling theory, 
the so-called schematic models, were found to be very 
useful in interpreting a variety of experimental data. These schematic
models were even used to analyze systems for which the original mode-coupling
theory was not intended, like molecular or polymeric fluids. Subsequently, 
the fully microscopic mode-coupling theory has been extended to treat some of 
these systems  \cite{Goetzebook}. 

In our opinion, the most valuable tests of the mode-coupling
theory were provided by computer simulations. The main reason for this is 
quite obvious; the same, well defined system can be used to derive 
theoretical predictions and to perform computer simulations. Thus, any disagreement
between theory and simulations reveals an inadequacy of the theory.
There is an additional reason for the the usefulness of computer simulation
tests of the mode-coupling theory. 
As we discuss below, this theory was found to 
describe slightly supercooled fluids. For the last
twenty years the region of applicability of the mode-coupling theory 
was easily accessible to computer simulation studies. For these two
reasons, in the following paragraphs we concentrate on the results
obtained by comparing predictions of the mode-coupling theory with 
results of computer simulations. We refer the reader to Refs. 
\cite{GoetzeJPCM1999,Goetzebook} for an extended comparison of 
theoretical and experimental results. 

Simulational tests \cite{KobLH} of the mode-coupling theory 
showed that it describes rather well the initial phase of the 
slowing down of the fluid's dynamics upon approaching the glass transition. 
In particular, the theory accounts for the so-called cage effect: in a fluid approaching
the glass transition a given particle spends considerable time in its solvation
shell before making any significant motion. This simple physical picture 
of a particle's motion is reflected
in a characteristic plateau in the mean-square displacement and in a two-step
decay of the so-called intermediate scattering functions. Mode-coupling theory's
predictions for these functions are in a good agreement with computer simulation
results. In particular, we shall mention here the accuracy of the theory's 
predictions for the intermediate time plateau of the scattering function, 
which is well approximated by the so-called critical 
non-ergodicity parameter \cite{NK,KNS}, and for the time dependence in the
plateau region, \textit{i.e.} the so-called $\beta$-scale relaxation 
\cite{GK2000,WPFV2010}.

The mode-coupling theory predicts that upon sufficient supercooling a fluid
undergoes an ergodicity breaking transition. Furthermore, the theory predicts that 
upon approaching this transition the relaxation time and the 
self-diffusion coefficient exhibit, respectively, a power law divergence
and a power law decay. Over approximately three decades of change 
of the relaxation time and self-diffusion coefficient, the latter 
predictions agree rather well with computer simulation results.
To be more precise, power laws can be fitted to computer simulation results 
and the resulting exponents are close to those predicted by the theory.
It has to be admitted, however, that the so-called mode-coupling temperature that is 
obtained from power law fits is usually quite different from the temperature of the 
ergodicity breaking transition predicted by the theory.
The difference is smaller for the so-called mode-coupling volume fraction
which is obtained from power law fits for hard sphere systems. 

The most important negative conclusion from computer simulation studies is that 
the ergodicity breaking transition predicted by the mode-coupling 
theory is absent. Thus, for strongly supercooled fluids theoretical predictions
and computer simulation results are completely different. 
Upon approaching the empirical mode-coupling transition point (\textit{i.e.}
the point determined by fitting
procedure mentioned above), there is a crossover regime in which one observes 
departures of computer simulation results for the relaxation time and the 
self-diffusion coefficient from mode-coupling power laws. 
It has to be noted that until recently, standard computer simulations 
(excluding Monte Carlo simulations utilizing specially devised, 
usually non-local, moves) 
could only \emph{approach} the above mentioned mode-coupling transition point. 
Even now, systematic studies of well equilibrated systems at and below the 
mode-coupling temperature (or at and above the mode-coupling volume fraction) 
are quite rare. 

Some of the reasons responsible for the controversy surrounding 
the mode-coupling theory have already been mentioned. Critics 
of the theory emphasize the fact that it can only describe the initial three
decades of the slowing down and that it predicts a spurious (non-existent) 
ergodicity breaking transition. Furthermore, they point out the discrepancy
between the mode-coupling temperature or volume fraction obtained from 
fitting the simulation results and the corresponding quantities predicted
by the theory.

In addition, about fifteen years ago, it was realized that there is a very interesting
phenomenon that accompanies the glass transition which cannot be 
described by the mode-coupling theory. A concerted simulational and experimental
effort revealed that upon approaching the glass transition dynamics not only slow down
but also become increasingly heterogeneous \cite{DHbook}. 
The so-called dynamic heterogeneities
can be quantified in terms of four-point correlation functions that describe
space and time-dependent correlations of the dynamics of individual particles. 
These correlation functions are very similar (although not identical) to 
the four-point function that is factorized in the standard derivation of 
the mode-coupling theory. One could argue that, since the mode-coupling theory
is based upon the factorization approximation, it necessarily 
neglects the existence of dynamic fluctuations,
it cannot describe dynamic heterogeneities, and thus constitutes a mean-field
theory of the glass transition.

We should recall at this point that within a standard static mean-field
theory there is an indirect way to calculate correlations (which in principle 
are neglected in the derivation of the mean-field equation of state). To this
end one introduces an external field and shows that a susceptibility 
describing the change of the order parameter due to the external field diverges
at the mean-field transition. Since the susceptibility can be easily related to 
a correlation function, in such a calculation one effectively uses a 
mean-field theory to reveal divergent fluctuations. 

The above described standard mean-field procedure was implemented by Biroli
\textit{et al.} \cite{BBMR} as an inhomogeneous mode-coupling theory. 
Specifically, Biroli \textit{et al.} calculated the so-called
three-point susceptibility that describes the change of the intermediate
scattering function (a two-point function) due to the presence 
of an external potential. They showed that the three-point
susceptibility diverges upon approaching the ergodicity breaking transition
of the mode-coupling theory.
In addition, it exhibits a divergent length upon approaching 
this transition. This behavior of the susceptibility 
is quite analogous to what is found in the mean-field calculation. 
The analogy is somewhat incomplete in that  
in the standard mean-field calculation one can easily relate the divergent
susceptibility to a divergent static correlation function. In contrast,
the relationship of the three-point susceptibility of the inhomogeneous 
mode-coupling theory to any correlation function is rather unclear
(and therefore a direct simulational test of inhomogeneous mode
coupling theory's predictions would require a rather difficult simulational
evaluation of the three-point susceptibility).
In spite of this fact, Biroli \textit{et al.}'s calculation suggests that 
the mode-coupling theory is indeed a mean-field theory of the glass transition.

On the other hand, results of recent mode-coupling calculations and computer 
simulations in higher spatial dimensions raised some doubts about the mean-field
character of the mode-coupling theory. The reason for this is that this
theory does not seem to become more accurate in higher spatial dimension, which
is a behavior that one would expect of a mean-field theory. 
First, it was showed \cite{SchmidSchillingPRE2010,IkedaMiyazakiPRL2011} 
that for hard spheres 
in high spatial dimensions the ergodicity breaking transition volume fraction 
predicted by the mode-coupling theory lies above the so-called dynamic
transition volume fraction and even above the Kauzmann transition volume fraction  
which are predicted by a static replica 
theory \cite{MezardParisi1999,ParisiZamponi2010}. 
Since the latter theory also aspires to be a mean-field theory of the glass transition, 
the difference between these predictions is rather disconcerting and suggests 
that at least one of these theories may be incorrect. Moreover, as pointed out
by Ikeda and Miyazaki \cite{IkedaMiyazakiPRL2011}, in higher spatial dimensions, 
the long time limit of the self part of the van Hove
function at the mode-coupling transition develops unphysical negative tails. 
Finally, results of recent 
computer simulations studies \cite{CIPZPRL2011} in higher spatial dimensions 
seem to be consistent with the replica approach and, therefore, suggest that
the mode-coupling theory might not be a correct mean-field theory of the glass 
transition. In our opinion more work is needed to fully resolve this issue. 

Somewhat surprisingly, during most of the thirty years of the existence of 
the mode-coupling theory, relatively 
little work has been done on the investigation of its most fundamental approximation,
\textit{i.e.} the factorization approximation,
and on the development of extensions and improvements of the theory. In our opinion
this was, in part, due to the original derivation of
the most widely applied version of the theory, which was 
reviewed in details in Refs. \cite{Goetzebook,GoetzeLH}. This derivation, 
while well suited to obtain rather quickly the mode-coupling equations, is an 
inconvenient starting point for calculating corrections to the standard theory. 
It is only relatively recently that several alternative, diagrammatic
derivations of the mode-coupling theory have been 
proposed \cite{Andreanovdiag,KimKawasakidiag,NishinoHayakawadiag}.
Notably, most of these derivations are quite complicated. Thus, it is not clear whether
they could be used to calculate corrections to the mode-coupling theory. 

We mention here two related but different attempts to derive 
extensions of the standard mode-coupling theory, which 
were proposed shortly after the original theory was 
derived. Das and Mazenko \cite{DM} 
showed that the sharp ergodic-nonergodic transition that Ref. \cite{DMRT} predicted is
cut off if, in addition to the mode-coupling diagrams, one also includes  
diagrams that enforce the standard relationship 
between the momentum density, the particle density and the velocity field.
At almost the same time G\"otze and Sj\"ogren \cite{GScutoff} showed
that the transition predicted by the version of the theory proposed in Ref. \cite{BGS}
is cut off due to coupling to current modes. Subsequently, it was argued that 
the latter cut off should be understood as a hopping or an activated process.

Recently, these two approaches, and related ones presented later in Refs. 
\cite{SDD,MazenkoYeoJSP1994}, 
were criticized by Cates and Ramaswamy \cite{CR}. 
These authors argued in quite general terms that couplings to current modes result
in negligible contributions and cannot induce hopping or activated processes.

We shall mention here that there is another reason why coupling to
current modes cannot constitute a universal extension of mode-coupling theory 
which cuts off the spurious transition and cures other problems
of this theory. The reason is that the long-time dynamics of systems in which
the underlying (microscopic) dynamics is Brownian is surprisingly similar to that of
systems evolving with Newtonian dynamics. It has been known for some time
that at the level of the standard mode-coupling approximation 
Brownian and Newtonian microscopic dynamics result in the same glass transition
scenario \cite{SL}. Later, it was showed using computer simulations 
that deviations from the mode-coupling-like behavior are the same in systems 
with stochastic dynamics and Newtonian dynamics \cite{GKB} and in systems 
with Brownian dynamics and Newtonian dynamics \cite{SE}. The implication of 
these studies is that the mechanism that cuts off the spurious transition
predicted by the mode-coupling theory is likely the same in systems with 
different microscopic dynamics. Since in systems with Brownian dynamics current
modes cannot be defined (at least not in the same way as in systems with 
Newtonian dynamics), the mechanism introduced in Refs. \cite{DM,GScutoff,SDD} cannot
operate there. 

Finally, we shall also mention here the so-called generalized mode-coupling approach.
This line of research was started when we recognized \cite{GSPRL2003} that by moving 
mode-coupling theory's factorization approximation to a higher level 
correlation function 
the location of the ergodicity breaking transition predicted by the theory
can be moved towards the empirical 
transition determined by fitting simulational data to mode-coupling-like
power laws. Subsequently, Wu and Cao  \cite{WuCaoPRL2005}
extended our calculation and showed that by 
moving the factorization approximation by two levels one can get even better 
agreement between theory and simulations. Finally, 
Mayer \textit{et al.} \cite{MayerMiyazakiReichmanPRL2006}
showed at the level of a schematic model that if one avoids the factorization
approximation altogether, the resulting theory does not have a spurious 
ergodicity breaking transition. On the one hand, this development looks quite 
promising. We showed  \cite{GSSendai}, however,  that  
from the diagrammatic point of view, the generalized mode-coupling
theory re-sums fewer diagrams than the the standard mode-coupling
theory. It is known in the liquid state theory that re-summing more
diagrams does not always result in a more accurate theory. It would, however, be
more satisfactory to correct mode-coupling approach by adding diagrams 
that describe dynamic events that are neglected in the standard mode-coupling approach.  

In the reminder of this paper we will review our diagrammatic 
formulation \cite{GSJCP2007} of the 
dynamics of strongly interacting systems of Brownian particles. We will show that 
this approach can be used to derive, in a rather straightforward way, 
the standard version of the 
mode-coupling theory. Finally, we will show that it can also be used to
incorporate dynamic events that are neglected in the standard theory. 
Specifically, we will evaluate the simplest corrections to the mode-coupling theory's
expression for the so-called irreducible memory function. 

\section{Diagrammatic approach}\label{sec:diagram}

\subsection{Derivation}\label{subsec:derivation} 

We consider a system of $N$ interacting Brownian particles in a volume $V$.
The average density is $n=N/V$. The brackets $\left< ... \right>$ 
indicate a canonical ensemble average at a temperature $T$. 
As shown in Ref.  \cite{GSJCP2007},
after some preliminary calculations it is convenient
to take the thermodynamic limit, $N\to\infty, V\to\infty, N/V=n=const$. 

We define the time dependent equilibrium correlation function of the Fourier components 
of the microscopic density as
\begin{equation}\label{dcfdef}
\left<n(\mathbf{k}_1;t) n^*(\mathbf{k}_2)\right>,
\end{equation}
with $n(\mathbf{k}_1;t)$ being the Fourier transform of the microscopic 
density fluctuation at a time $t$, 
\begin{equation}\label{denfldef}
n(\mathbf{k}_1;t) = \sum_{j=1}^N
e^{-i\mathbf{k}_1\cdot\mathbf{r}_j(t)} - \left<\sum_{j=1}^N
e^{-i\mathbf{k}_1\cdot\mathbf{r}_j}\right>,
\end{equation}
and $n(\mathbf{k}_2) \equiv n(\mathbf{k}_2;t=0)$. 
In a diagrammatic
series it is convenient to express the density
correlation function in terms of the so-called response function $G(k;t)$, 
\begin{equation}\label{Gdef}
\theta(t) \left<n_1(\mathbf{k}_1;t)n_1^*(\mathbf{k}_2) \right> = 
n G(k;t) S(k) (2\pi)^3 \delta(\mathbf{k}_1-\mathbf{k}_2).
\end{equation} 
Note that due to the translational invariance, the correlation function 
$\left<n_1(\mathbf{k}_1;t)n_1^*(\mathbf{k}_2) \right>$
is diagonal in wave-vector space. 
The response function is related to the usual collective  
intermediate scattering function $F(k;t)$,
\begin{equation}\label{GFconnection}
F(k;t) = G(k;t) S(k).
\end{equation}
To derive the diagrammatic series for the response function $G$ it is convenient to start
from a hierarchy of equations of motion for the correlation functions 
of orthogonalized densities. The first such correlation function coincides with 
formula (\ref{dcfdef}). The second one,
\begin{equation}\label{2dcfdef}
\left<n_2(\mathbf{k}_1,\mathbf{k}_2;t) n^*(\mathbf{k}_3)\right>,
\end{equation}
is a correlation function of the 
part of the two-particle density fluctuation that is orthogonal to the density
fluctuation, $n_2(\mathbf{k}_1,\mathbf{k}_2;t)$, 
\begin{eqnarray}\label{denf2def}
n_2(\mathbf{k}_1,\mathbf{k}_2;t) &=& \sum_{i\neq j =1}^N
e^{-i\mathbf{k}_1\cdot\mathbf{r}_i(t)-i\mathbf{k}_2\cdot\mathbf{r}_j(t)} 
- \left< \sum_{i\neq j =1}^N
e^{-i\mathbf{k}_1\cdot\mathbf{r}_i-i\mathbf{k}_2\cdot\mathbf{r}_j} \right> 
\nonumber \\ && 
- \sum_{\mathbf{q}_1,\mathbf{q}_2} 
\left< \sum_{i\neq j =1}^N
e^{-i\mathbf{k}_1\cdot\mathbf{r}_i-i\mathbf{k}_2\cdot\mathbf{r}_j}
n^*(\mathbf{q}_1)\right> \left< n(\mathbf{q}_1) n^*(\mathbf{q}_2)\right>^{-1}
n(\mathbf{q}_2;t) 
, \nonumber \\
\end{eqnarray}
and the density fluctuation $ n(\mathbf{k}_3)$. One should note that by
definition $\left<n_2(\mathbf{k}_1,\mathbf{k}_2;t=0) n^*(\mathbf{k}_3)\right>=0$.

We shall mention here that 
the orthogonalized densities were introduced before \cite{H1,H2} in 
the context of a diagrammatic approach to the dynamics of Newtonian systems. 
The advantage of describing the system in terms of correlation functions of 
orthogonalized densities is two-fold. First, the introduction of the orthogonalized
densities allows us to avoid having additional, rather unusual, diagrams that 
impose the equilibrium distribution at the initial time. Technically, this 
follows from the vanishing of all higher order correlation at $t=0$,
$\left<n_l(\mathbf{k}_1,...,\mathbf{k}_l;t=0) n^*(\mathbf{k}_{l+1})\right>=0$.
Second, if equations of
motion are written in terms of the correlation functions of 
orthogonalized densities, the bare
inter-particle interactions are automatically renormalized. Specifically, 
in the equations of motions the inter-particle potential is
replaced by combinations of equilibrium correlation functions. The disadvantage
of using orthogonalized densities is that in addition to equilibrium pair correlations,
many-particle correlations are needed to express renormalized interactions. 
To make our approach tractable we perform a cluster expansion of
the renormalized interactions and neglect terms involving higher order
equilibrium correlations. While this approximation is implicit in all 
recent diagrammatic approaches to the dynamics of strongly interacting fluids,
its consequences have yet to be investigated. 

As shown in Ref. \cite{GSJCP2007}, 
the hierarchy of equations of motion for the correlation functions 
of orthogonalized densities can be replaced by a hierarchy of integral
equations involving the same functions. The latter hierarchy can be solved by iteration
for the response function (\ref{Gdef}) 
and the resulting expressions can be represented in terms of diagrams. 
The diagrams consist of the following elements:
\begin{itemize}
\item response function $G(k;t)$:
\includegraphics[scale=.17]{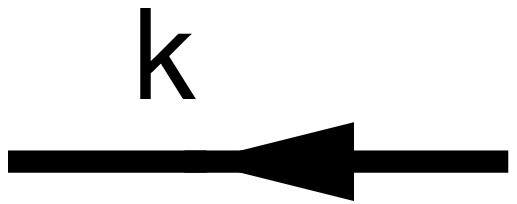}
\item bare response function $G_0(k;t)$:
\includegraphics[scale=.17]{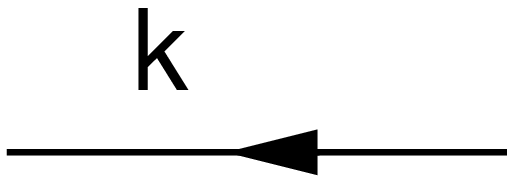}
\item ``left'' vertex $\mathcal{V}_{12}$:
\includegraphics[scale=.17]{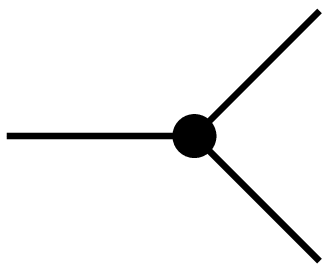}
\item ``right'' vertex $\mathcal{V}_{21}$: 
\includegraphics[scale=.17]{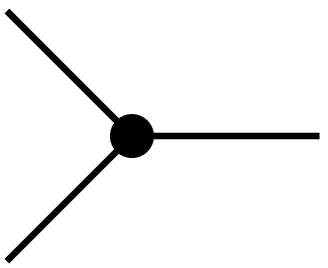}
\item four-leg vertex $\mathcal{V}_{22}$:
\includegraphics[scale=.17]{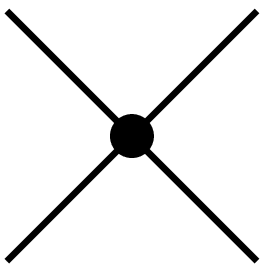}
\end{itemize}
The bare response function $G_0(k;t)$ is defined as
\begin{equation}\label{G0def}
G_0(k;t) = \theta(t) \exp(-D_0 k^2 t/S(k)),
\end{equation}
and the explicit expressions for the three- and four-leg vertices are:
\begin{eqnarray}\label{V12}
\mathcal{V}_{12}(\mathbf{k}_1;\mathbf{k}_2,\mathbf{k}_3) = 
D_0 (2\pi)^3 \delta(\mathbf{k}_1-\mathbf{k}_2-\mathbf{k}_3)
k_1 v_{\mathbf{k}_1}(\mathbf{k}_2,\mathbf{k}_3)
\end{eqnarray}
\begin{eqnarray}\label{V21}
\mathcal{V}_{21}(\mathbf{k}_1,\mathbf{k}_2;\mathbf{k}_3) =
n D_0 (2\pi)^3 \delta(\mathbf{k}_1+\mathbf{k}_2-\mathbf{k}_3)S(k_1)S(k_2)
k_3 v_{\mathbf{k}_3}(\mathbf{k}_1,\mathbf{k}_1)
S(k_3)^{-1},
\end{eqnarray}
\begin{eqnarray}\label{V22}
\lefteqn{
\mathcal{V}_{22}(\mathbf{k}_1,\mathbf{k}_2;\mathbf{k}_3,\mathbf{k}_4) = } 
\\ \nonumber &&
n D_0 (2\pi)^3 S(k_1)S(k_2)\delta(\mathbf{k}_1+\mathbf{k}_2-\mathbf{k}_3-\mathbf{k}_4)
\mathbf{v}(\mathbf{k}_1,\mathbf{k}_2)\cdot\mathbf{v}(\mathbf{k}_3,\mathbf{k}_4)
\end{eqnarray}
In Eqs. (\ref{V12}-\ref{V21}), vertices 
$\mathcal{V}_{12}$ and $\mathcal{V}_{21}$ are expressed in terms of the
following function, 
\begin{eqnarray}\label{vdef} 
v_{\mathbf{k}_1}(\mathbf{k}_2,\mathbf{k}_3)
= \hat{\mathbf{k}}_1\cdot \left(c(k_2)\mathbf{k}_2+c(k_3)\mathbf{k}_3\right).
\end{eqnarray}
In the literature, $v_{\mathbf{k}_1}(\mathbf{k}_2,\mathbf{k}_3)$ 
is referred to as the vertex function of the mode-coupling theory. 
Furthermore, in Eq. (\ref{V22}), vertex $\mathcal{V}_{22}$ is expressed in
terms of a similar function,
\begin{eqnarray}\label{vbdef} 
\mathbf{v}(\mathbf{k}_1,\mathbf{k}_2)=
c(k_1)\mathbf{k}_1+c(k_2)\mathbf{k}_2.
\end{eqnarray}
In later sections we will also use the following functions related to 
$v$ and $\mathbf{v}$,
\begin{eqnarray}\label{vdef1} 
\tilde{v}_{\mathbf{k}_1}(\mathbf{k}_2,\mathbf{k}_3)
= \hat{\mathbf{k}}_1\cdot \left(c(k_2)\mathbf{k}_2+c(k_3)\mathbf{k}_3\right)/k_1,
\end{eqnarray}
\begin{eqnarray}\label{vbdef1} 
\tilde{\mathbf{v}}(\mathbf{k}_1,\mathbf{k}_2)=
\left(c(k_1)\mathbf{k}_1+c(k_2)\mathbf{k}_2\right)/|\mathbf{k}_1+\mathbf{k}_2|.
\end{eqnarray}

In the diagrams contributing to the response function, 
we refer to the leftmost bare response function as the 
left root, and to the other bare response functions as bonds. The left root
is labeled by a wave-vector and the bonds are unlabeled. 
We consider two diagrams to be topologically
equivalent if there is a way to assign labels to unlabeled bonds so that the
resulting labeled diagrams are topologically equivalent\footnote{Two labeled diagrams
are topologically equivalent if each labeled bond in one diagram connects 
vertices of the same type as the corresponding labeled bond in the other 
diagram  \cite{Mayer}.}. To evaluate an unlabeled
diagram one assigns wave-vectors to unlabeled bonds, 
integrates over all wave-vectors (with a $(2\pi)^{-3}$ 
factor for each integration) except the wave-vector
corresponding to the left root, integrates over all intermediate times, 
and divides the result by a symmetry number of the
diagram (\textit{i.e.} the number of topologically identical labeled diagrams
that can be obtained from a given unlabeled diagram by permutation of
the bond labels). 

\begin{figure}
\centerline{\includegraphics[scale=.17]{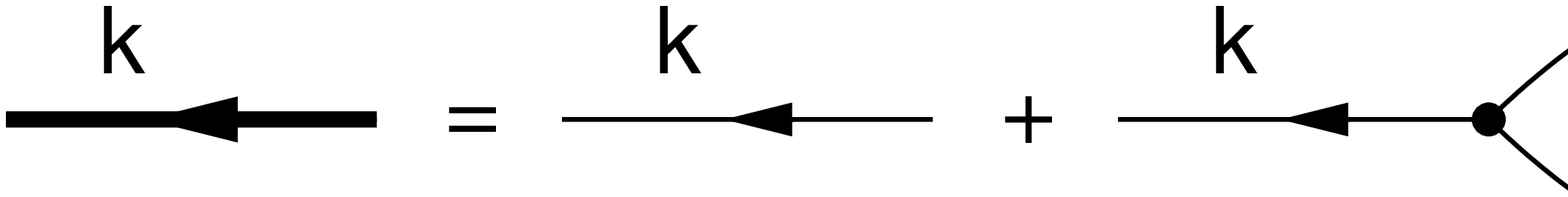}}
\caption{Diagrammatic series expansion for 
response function $G(k;t)$  \cite{GSJCP2007}.} 
\label{f:Giter}
\end{figure}

As showed in Ref. \cite{GSJCP2007}, the response function given by the following series:
\begin{eqnarray}\label{Gseries}
&& G(k;t) = \\ &&  \nonumber 
\mbox{sum of all topologically different 
diagrams with a left root labeled $k$,  }  \\ && \nonumber 
\mbox{a right root, 
$G_0$ bonds, $\mathcal{V}_{12}$, $\mathcal{V}_{21}$ and $\mathcal{V}_{22}$
vertices, in which diagrams}  \\ &&  \nonumber
\mbox{with odd and even numbers of 
$\mathcal{V}_{22}$ vertices contribute with overall }  \\ &&  \nonumber
\mbox{negative and positive sign, respectively.}
\end{eqnarray}
The first few diagrams contributing to the series (\ref{Gseries}) 
are shown in Fig. \ref{f:Giter}.

\subsection{Memory functions: reducible and irreducible}\label{subsec:mf} 

We should emphasize that our diagrammatic expansion was not derived from 
a field-theoretical approach. However, once a diagrammatic approach has been derived, 
we can use re-arrangements and re-summations that were originally introduced in
the context of field-theoretical diagrammatic expansions. 
In particular, we can write down a Dyson equation in the usual form,
\begin{eqnarray}\label{Dyson}
G(k;t) = G_0(k;t) + \int_0^t dt_1 \int_0^{t_1} dt_2 \int \frac{d\mathbf{k}_1}{(2\pi)^3} 
G_0(k;t-t_1)  \Sigma(\mathbf{k},\mathbf{k}_1;t_1-t_2) G(k_1;t_2). 
\nonumber \\
\end{eqnarray}

\begin{figure}
\centerline{\includegraphics[scale=.17]{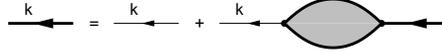}}
\caption{Diagrammatic representation of the Dyson equation, Eq. (\ref{Dyson})
\cite{GSJCP2007}.}
\label{f:dyson}
\end{figure}

Here $\Sigma$ is the self-energy.
The diagrammatic representation of the Dyson equation is showed in Fig. \ref{f:dyson}.
Due to the translational invariance the self-energy is diagonal in the wave-vector space,
\begin{equation}\label{sediag}
\Sigma(\mathbf{k},\mathbf{k}_1;t) \propto (2\pi)^3 \delta(\mathbf{k}-\mathbf{k}_1).
\end{equation}

It can be showed from the Dyson equation that 
the self-energy $\Sigma$ is a sum of diagrams that do not separate 
into disconnected components upon removal of a single bond. To make 
a connection with the projection operator-based approach \cite{SL,HK} we need
to relate the self-energy to a memory function. First, we
note that the diagrams contributing to the self-energy
start with $\mathcal{V}_{21}$ vertex on the right and end with 
$\mathcal{V}_{12}$ vertex on the left. It turns out that in order to relate the
self-energy to a memory function
for a Brownian system, we need to factor out parts of these vertices. First, we define a
memory matrix $\mathbf{M}$ by factoring out $\mathbf{k}$ from the left  vertex
and $(D_0/S(k_1)) \mathbf{k}_1$ from the right vertex,
\begin{equation}\label{matm}
\Sigma(\mathbf{k},\mathbf{k}_1;t) = 
D_0
\mathbf{k}\cdot\mathbf{M}(\mathbf{k},\mathbf{k}_1;t)\cdot\mathbf{k}_1 S(k_1)^{-1}.
\end{equation}
Due to translational invariance the memory matrix $\mathbf{M}$ is
diagonal in the wave-vector space. Moreover, only its longitudinal component contributes
to the self-energy. Thus, we can define the memory
function $M$ through the following relation
\begin{equation}\label{matmdiag}
\hat{\mathbf{k}}\cdot\mathbf{M}(\mathbf{k},\mathbf{k}_1;t)\cdot\hat{\mathbf{k}} 
= M(k;t) (2\pi)^3 \delta(\mathbf{k}-\mathbf{k}_1).
\end{equation}
Using Eqs. (\ref{matm}) and (\ref{matmdiag}) in the Laplace
transform of the Dyson equation, we can obtain the following equation,
\begin{equation}\label{Dysonm}
G(k;z) = G_0(k;z) + G_0(k;z) \frac{D_0 k^2}{S(k)} M(k;z) G(k;z).
\end{equation}
Eq. (\ref{Dysonm}) can be solved with respect to (w.r.t.) response function $G(k;z)$.
Using the definition of bare response function $G_0$ we obtain 
\begin{equation}\label{Gm}
G(k;z) = \frac{1}{z+ \frac{D_0 k^2}{S(k)} \left(1-M(k;z)\right)}.
\end{equation}
Multiplying both sides of the above equation by the static structure
factor and using the relation (\ref{GFconnection}) between $G$ and the
intermediate scattering function $F$ we get the standard 
memory function representation  \cite{HK} of the intermediate scattering function,
\begin{equation}\label{Fm}
F(k;z) = \frac{S(k)}{z+ \frac{D_0 k^2}{S(k)} \left(1-M(k;z)\right)}.
\end{equation}
The memory function representation (\ref{Fm}) is the first step in the 
derivation of the mode-coupling equations that utilizes the projection operator
formalism. 

To analyze the diagrams contributing to the memory matrix it is convenient to 
introduce cut-out vertices:
\begin{eqnarray} \label{V12c}
\mathbf{V}_{12}^{\mathrm{c}}(\mathbf{k}_1;\mathbf{k}_2,\mathbf{k}_3) =
D_0 (2\pi)^3 \delta(\mathbf{k}_1-\mathbf{k}_2-\mathbf{k}_3) 
\left(c(k_2)\mathbf{k}_2+c(k_3)\mathbf{k}_3\right)
\end{eqnarray}
\begin{eqnarray}\label{V21c}
\mathbf{V}_{21}^{\mathrm{c}}(\mathbf{k}_1,\mathbf{k}_2;\mathbf{k}_3) =
n (2\pi)^3 \delta(\mathbf{k}_1+\mathbf{k}_2-\mathbf{k}_3)S(k_1)S(k_2)
\left(c(k_1)\mathbf{k}_1+c(k_2)\mathbf{k}_2\right).
\end{eqnarray}
These vertices are obtained by factoring out $\mathbf{k}_1$ from 
vertex $\mathcal{V}_{12}$ and $(D_0/S(k_3)) \mathbf{k}_3$
from vertex $\mathcal{V}_{21}$ (one should note that the same factorization
was used in the definition of the memory matrix in Eq. (\ref{matm})).

It should be noted that 
\begin{eqnarray}\label{fact}
\mathcal{V}_{22}(\mathbf{k}_1,\mathbf{k}_2;\mathbf{k}_3,\mathbf{k}_4)  
= \int \frac{d \mathbf{k}'}{(2\pi)^3}
\mathbf{V}_{21}^{\mathrm{c}}(\mathbf{k}_1,\mathbf{k}_2;\mathbf{k}')\cdot
\mathbf{V}_{12}^{\mathrm{c}}(\mathbf{k}';\mathbf{k}_3,\mathbf{k}_4).
\end{eqnarray}
The diagrammatic 
rules for functions $\mathbf{V}_{12}^{\mathrm{c}}$ 
and $\mathbf{V}_{21}^{\mathrm{c}}$ are as follows:
\begin{itemize}
\item ``left'' cut-out vertex $\mathbf{V}_{12}^{\mathrm{c}}$:
\includegraphics[scale=.17]{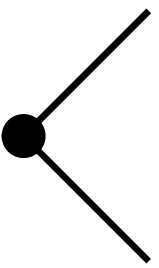}
\item ``right'' cut-out vertex $\mathbf{V}_{21}^{\mathrm{c}}$: 
\includegraphics[scale=.17]{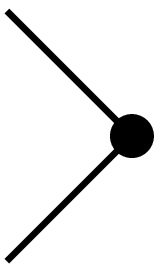}
\end{itemize}
and we refer to wave-vector $\mathbf{k}_1$ in 
$\mathbf{V}_{12}^{\mathrm{c}}(\mathbf{k}_1;\mathbf{k}_2,\mathbf{k}_3)$
and $\mathbf{k}_3$ in 
$\mathbf{V}_{21}^{\mathrm{c}}(\mathbf{k}_1,\mathbf{k}_2;\mathbf{k}_3)$
as roots of these vertices. Note that to evaluate a diagram contributing 
to the memory matrix we do not integrate over either left or right roots.

\begin{figure}
\centerline{\includegraphics[scale=.17]{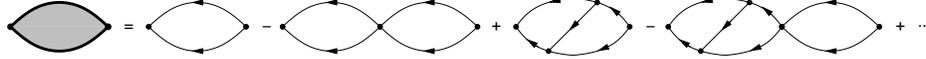}}
\caption{Diagrammatic series expansion for memory matrix $\mathbf{M}$.}
\label{f:mred}
\end{figure}

It follows from the definition of the memory matrix $\mathbf{M}$
that 
\begin{eqnarray}\label{Mredseries}
&& \mathbf{M}(\mathbf{k},\mathbf{k}_1;t) = \\ &&  \nonumber 
\mbox{sum of all topologically different diagrams which do not separate 
into}  \\ && \nonumber \mbox{disconnected 
components upon removal of a single bond, with vertex $\mathbf{V}_{12}^{\mathrm{c}}$}  
\\ && \nonumber 
\mbox
{with root $\mathbf{k}$ on the left and vertex $\mathbf{V}_{21}^{\mathrm{c}}$
with root $\mathbf{k}_1$ on the right,
$G_0$ bonds,} 
\\ && \nonumber  \mbox{$\mathcal{V}_{12}$, $\mathcal{V}_{21}$ and $\mathcal{V}_{22}$
vertices, in which diagrams with odd and even numbers of} 
\\ && \nonumber \mbox{$\mathcal{V}_{22}$ vertices contribute with
overall negative and positive sign, respectively.}
\end{eqnarray}
The first few diagrams in the series for $\mathbf{M}$ are showed in Fig. \ref{f:mred}.

The series expansion for $\mathbf{M}$ consists of diagrams that are
one-propagator irreducible (\textit{i.e.} diagrams that do not separate into 
disconnected components upon removal of a single bond). However, 
not all of these diagrams are completely one-particle irreducible. Specifically,
some of the diagrams contributing to $\mathbf{M}$ separate into disconnected
components upon removal of a $\mathcal{V}_{22}$ vertex. In analogy with
terminology used in the context of Mayer diagrams, we will call such a vertex
an articulation $\mathcal{V}_{22}$ vertex or articulation four-leg vertex.
The examples of diagrams containing one articulation $\mathcal{V}_{22}$ vertex 
are the second and
the fourth diagrams on the right-hand side of the diagrammatic equation
showed in Fig. \ref{f:mred}.
Intuitively, it is clear that the series (\ref{Mredseries}) can be further
re-arranged by writing down a second Dyson-type equation. In the  projection operator 
formalism, this second Dyson-type equation corresponds to the equation
that defines the so-called irreducible memory function  \cite{SL,CHess,Kawasaki}
in terms of the memory function defined through Eq. (\ref{matmdiag}). 

In the diagrammatic approach, 
we define the irreducible memory matrix $\mathbf{M}^{\mathrm{irr}}$ 
as a sum of only those diagrams
in the series for $\mathbf{M}$ that do not separate into disconnected components
upon removal of a single $\mathcal{V}_{22}$ vertex. 
To distinguish  memory matrix $\mathbf{M}$ from the irreducible matrix
$\mathbf{M}^{\mathrm{irr}}$ we will sometimes use the term reducible 
memory matrix when referring to $\mathbf{M}$. We will also sometimes us the 
term reducible memory function when referring to $M$ defined in Eq. (\ref{matmdiag}).

\begin{figure}
\centerline{\includegraphics[scale=.17]{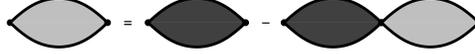}}
\caption{Memory matrix $\mathbf{M}$ can be represented as a sum of
$\mathbf{M}^{\mathrm{irr}}$ and all other diagrams  \cite{GSJCP2007}.
The latter diagrams can be re-summed and
it is easy to see that as a result we get the second diagram at the right-hand side.}
\label{f:mredirr}
\end{figure}

Diagrammatically, 
we can represent memory matrix $\mathbf{M}$ as a sum of
$\mathbf{M}^{\mathrm{irr}}$ and all other diagrams. The latter diagrams can
be re-summed as showed in Fig. \ref{f:mredirr}. 
Using Eq. (\ref{fact}), we can introduce an additional integration over a
wave-vector and then we see that the diagrammatic equation showed in Fig. \ref{f:mredirr}
corresponds to the following equation,
\begin{eqnarray}\label{mredirr} 
\mathbf{M}(\mathbf{k},\mathbf{k}_1;t)
= \mathbf{M}^{\mathrm{irr}}(\mathbf{k},\mathbf{k}_1;t)
- \int_0^t dt_1 \int \frac{d\mathbf{k}_2}{(2\pi)^3}
\mathbf{M}^{\mathrm{irr}}(\mathbf{k},\mathbf{k}_2;t-t_1)\cdot
\mathbf{M}(\mathbf{k}_2,\mathbf{k}_1;t_1)
\nonumber \\ 
\end{eqnarray}
Again, we use translational invariance and then introduce the 
irreducible memory function $M^{\mathrm{irr}}$ as the longitudinal component 
of the matrix $\mathbf{M}^{\mathrm{irr}}$, 
\begin{equation}\label{mirr}
\hat{\mathbf{k}}\cdot\mathbf{M}^{\mathrm{irr}}(\mathbf{k},\mathbf{k}_1;t)
\cdot\hat{\mathbf{k}} = 
M^{\mathrm{irr}}(k;t) (2\pi)^3 \delta(\mathbf{k}-\mathbf{k}_1),
\end{equation}
Then, the longitudinal component of the  Laplace transform of Eq. (\ref{mredirr})
can be written in the following way 
\begin{equation}\label{mredirr2}
M(k;z) = M^{\mathrm{irr}}(k;z) - M^{\mathrm{irr}}(k;z) M(k;z).
\end{equation}
This equation can be solved w.r.t. memory function $M$.  Substituting the solution
into Eq. (\ref{Fm}) we obtain a representation of the intermediate
scattering function in terms of the irreducible memory function,
\begin{equation}\label{Fmirr}
F(k;z) = S(k) G(k;z) = 
\frac{S(k)}{z+ \frac{D_0 k^2}{S(k)\left(1+M^{\mathrm{irr}}(k;z)\right)}}.
\end{equation}
Eq. (\ref{Fmirr}) was first derived by Cichocki and Hess \cite{CHess}
using a projection operator approach. Subsequently, it was used by Szamel and 
L\"owen \cite{SL} to derive the standard mode-coupling theory for Brownian systems.

Diagrammatically, 
\begin{eqnarray}\label{Mirrseries}
&& \mathbf{M}^{\mathrm{irr}}(\mathbf{k},\mathbf{k}_1;t) = \\ &&  \nonumber 
\mbox{sum of all topologically different diagrams which do not separate into}  
\\ && \nonumber 
\mbox{disconnected components upon removal of a single bond or a single 
$\mathcal{V}_{22}$}  \\ && \nonumber 
\mbox{vertex, with vertex $\mathbf{V}_{12}^{\mathrm{c}}$ 
with root $\mathbf{k}$ on the left and vertex $\mathbf{V}_{21}^{\mathrm{c}}$
with root $\mathbf{k}_1$}  \\ && \nonumber 
\mbox{on the right, $G_0$ bonds, $\mathcal{V}_{12}$, 
$\mathcal{V}_{21}$ and $\mathcal{V}_{22}$, in which diagrams with odd and}  
\\ && \nonumber 
\mbox{even numbers of $\mathcal{V}_{22}$ vertices contribute with 
overall negative and positive sign,}  \\ && \nonumber \mbox{respectively.}
\end{eqnarray}

\begin{figure}
\centerline{\includegraphics[scale=.17]{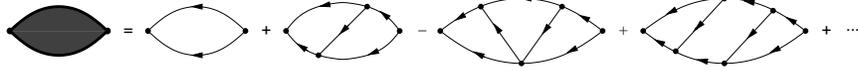}}
\caption{Diagrammatic series expansion for the irreducible 
memory matrix $\mathbf{M}^{\mathrm{irr}}$.}
\label{f:mirr}
\end{figure}

The first few diagrams in the series for $\mathbf{M}^{\mathrm{irr}}$ are shown in 
Fig. \ref{f:mirr}. We will analyze three classes of these diagrams 
in the following sections. Here we will 
only notice that the first diagram at the right-hand-side of the
diagrammatic equation showed in Fig. \ref{f:mirr} separates into two disconnected
pieces upon removal of the left and right vertices. The remaining three
diagrams do not share this property. We shall point out the important difference
between the second and third diagrams and the fourth diagram. 
The latter diagram is two-line-reducible, 
\textit{i.e.} it separates into two disconnected pieces 
upon removing the left and right vertices and 
cutting through two propagator lines (note that each of these pieces contains
at least two horizontal lines and is itself internally connected). 
Roughly speaking, the fourth diagram has the the same nontrivial part as the second 
diagram but iterated twice. In contrast, the second and third 
diagrams are two-line-irreducible: upon removing the left and right vertices
they cannot be separated into two internally connected pieces by cutting
through two propagator lines. 

\section{Standard mode-coupling approximation}\label{sec:mct}

To obtain the standard mode-coupling expression for the memory function it is convenient 
to start from a series expression for 
$\mathbf{M}^{\mathrm{irr}}$ showed in Fig. \ref{f:mirr}. 
The simplest re-summation of this series includes diagrams
that separate into two disconnected components upon removal of the left,
$\mathbf{V}_{12}^{\mathrm{c}}$, and the right, 
$\mathbf{V}_{21}^{\mathrm{c}}$, vertices. Out of the diagrams at the right-hand-side 
of the diagrammatic equation showed in Fig. \ref{f:mirr},
this re-summation includes only the first diagram. We will call 
diagrams that separate into two disconnected components upon removal of the left
and right vertices mode-coupling diagrams. 

\begin{figure}
\centerline{\includegraphics[scale=.17]{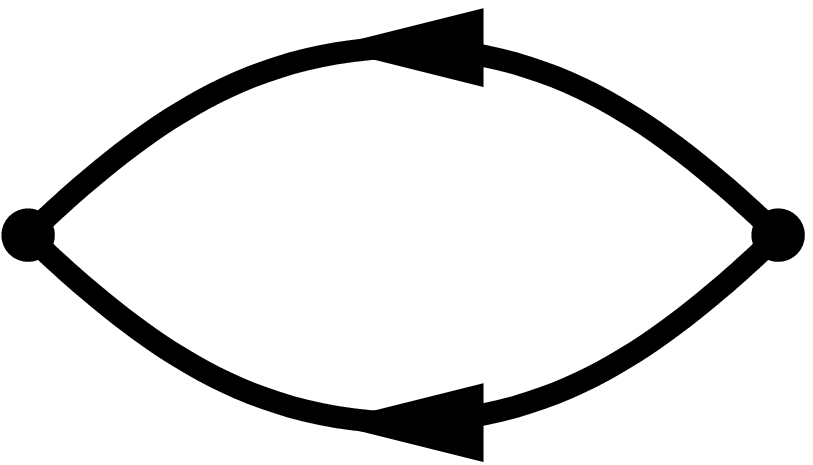}}
\caption{Re-summation of diagrams that separate into two disconnected 
components upon removal of 
the $\mathbf{V}_{12}^{\mathrm{c}}$ and $\mathbf{V}_{21}^{\mathrm{c}}$ vertices
leads to a one-loop diagram with $G$ bonds  \cite{GSJCP2007}.}
\label{f:mct}
\end{figure}

In the diagrams included in the present re-summation, 
each of the two components that appear after removing the left and right vertices
is a part of the series for the
response function $G$. Thus, the present  
re-summation results in a one-loop diagram 
(\textit{i.e.} the first diagram shown on the right-hand side in Fig. \ref{f:mirr}) 
with bare $G_0$ bonds replaced by $G$ bonds, see Fig. \ref{f:mct}.
Thus, we get a self-consistent one-loop approximation
for the memory matrix, 
\begin{eqnarray}\label{mct}
\lefteqn{\mathbf{M}^{\mathrm{irr}}(\mathbf{k},\mathbf{k}_1;t) \approx 
\mathbf{M}^{\mathrm{irr}}_{\mathrm{one-loop}}(\mathbf{k},\mathbf{k}_1;t) = } 
\\ \nonumber && 
\frac{1}{2} \int \frac{d\mathbf{k}_2 d\mathbf{k}_3}{(2\pi)^6} 
\mathbf{V}_{12}^{\mathrm{c}}(\mathbf{k};\mathbf{k}_2,\mathbf{k}_3) 
G(k_2;t) G(k_3;t) \mathbf{V}_{21}^{\mathrm{c}}(\mathbf{k}_2,\mathbf{k}_3;\mathbf{k}_1).
\end{eqnarray}
The overall factor $1/2$ reflects the symmetry number of the one-loop diagram,
which is equal to 2.

Using explicit expressions (\ref{V12c}-\ref{V21c}) for the cut-out vertices
we show that (\ref{mct}) leads to the following expression for the 
irreducible memory function
(recall that the irreducible memory function is obtained from 
the memory matrix by using translational invariance and
taking the matrix's longitudinal component):
\begin{eqnarray}\label{mctexplicit}
\lefteqn{M^{\mathrm{irr}}(k;t) \approx M^{\mathrm{irr}}_{\mathrm{one-loop}}(k;t) = } 
\\ && 
\frac{n D_0}{2} \int \frac{d\mathbf{k}_1}{(2\pi)^3} 
v_{\mathbf{k}}^2(\mathbf{k}_1,\mathbf{k}-\mathbf{k}_1)
S(k_1)S(|\mathbf{k}-\mathbf{k}_1|)G(k_1;t) G(|\mathbf{k}-\mathbf{k}_1|;t)
\equiv M^{\mathrm{irr}}_{\mathrm{MCT}}(k;t)
\nonumber
\end{eqnarray}
where $v_{\mathbf{k}}(\mathbf{k}_1,\mathbf{k}-\mathbf{k}_1)$ 
denotes the so-called mode-coupling theory's vertex defined in Eq. (\ref{vdef}).
As indicated in Eq. (\ref{mctexplicit}), 
the self-consistent one-loop approximation coincides with
the standard mode-coupling approximation, \textit{i.e.} both approximations result
in exactly the same expression for the irreducible memory function.
By combining the 
memory function representation (\ref{Fmirr}) with the standard mode-coupling 
approximation for the memory function (\ref{mctexplicit}), one can derive existence
and analyze the properties of an ergodicity breaking transition. 

More generally, 
if one assumes that at a certain state point the response function 
acquires a non-vanishing long-time limit,
\begin{eqnarray}\label{mctltl}
\lim_{t\to\infty} G(k;t) = f(k),
\end{eqnarray}
where $f$ is referred to as an non-ergodicity parameter, 
using Eqs. (\ref{Fmirr}) one can derive the 
well know equation for $f$,
\begin{eqnarray}\label{mctnep}
\frac{f(k)}{1-f(k)} = m(k).
\end{eqnarray}
In Eq. (\ref{mctnep}), $m(k)$ is related to the long-time limit of the  
irreducible memory function,
\begin{eqnarray}\label{mdef}
m(k) = \lim_{t\to\infty} \frac{S(k)}{D_0 k^2} M^{\mathrm{irr}}(k;t).
\end{eqnarray}
It should be emphasized that Eqs. (\ref{mctnep}-\ref{mdef}) are independent
of the mode-coupling approximation and, in fact, are exact. 
Specifically, if the response function does not decay, its long-time limit
is connected to the long-time limit of the memory function via 
Eqs. (\ref{mctnep}-\ref{mdef}).

Within the standard mode-coupling approximation $m$ is given by
\begin{eqnarray}\label{mmctdef}
&& m(k) \approx m_{\mathrm{MCT}}(k) = 
\\ \nonumber && 
\frac{n S(k)}{2} \int \frac{d\mathbf{k}_1}{(2\pi)^3} 
\tilde{v}_{\mathbf{k}}^2(\mathbf{k}_1,\mathbf{k}-\mathbf{k}_1)
S(k_1)S(|\mathbf{k}-\mathbf{k}_1|)f(k_1) f(|\mathbf{k}-\mathbf{k}_1|),
\end{eqnarray}
where we used a modified vertex function 
$\tilde{v}$ defined in Eq. (\ref{vdef1}).

\section{Two corrections to the standard mode-coupling
approximation}\label{sec:beyond}

\subsection{General considerations}\label{subsec:beyond0}

To improve upon the standard mode-coupling approximation we need to 
include at least some of the diagrams that are neglected in the re-summation
leading to the self-consistent one-loop approximation for the memory matrix.
For example, we might include some or all of 2nd, 3rd, or 4th diagrams 
at the right-hand-side of the 
diagrammatic equation showed in Fig. \ref{f:mirr}. We will refer to such 
diagrams, \textit{i.e.} to diagrams contributing to the irreducible memory
matrix which do not separate into disconnected components upon removing the
left and right vertices, as non-mode-coupling diagrams. 

\begin{figure}
\centerline{\includegraphics[scale=.17]{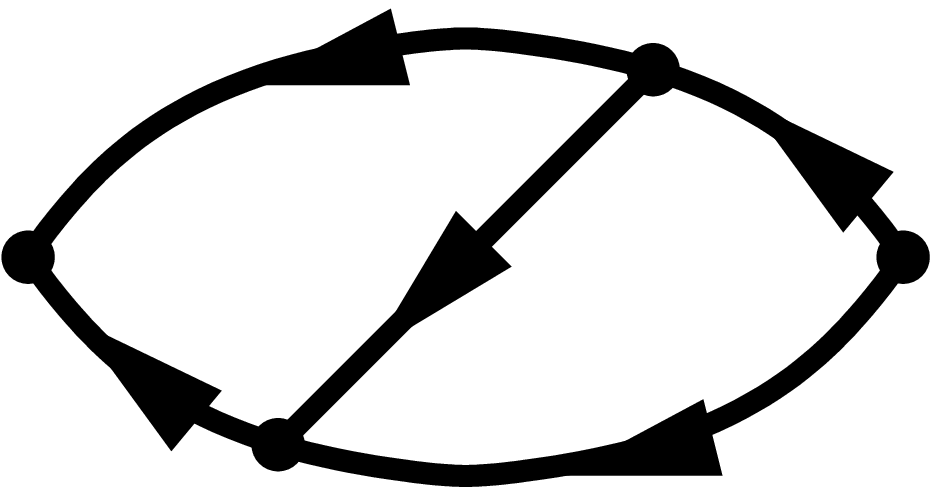}}
\caption{Re-summation of diagrams that can be obtained from
the 2nd diagram at the right-hand-side in Fig. \ref{f:mirr} by replacing the
bare response functions $G_0$ by diagrams that appear in the diagrammatic
expansion for the full response function, Eq. (\ref{Gseries}), gives 
the same diagram but with $G$ bonds.}
\label{f:bmct1}
\end{figure}

The simplest non-mode-coupling diagram is the 2nd diagram showed 
at the right-hand-side of the diagrammatic equation in Fig. \ref{f:mirr}.
Of course, including just the ``bare'' 2nd diagram (\textit{i.e} including
the 2nd diagram with $G_0$ bonds) would only introduce a trivial change
of the irreducible memory function. 
Instead, one should try to perform a re-summation of the diagrams 
with the same ``skeleton'' as the 2nd diagram at the right-hand-side in 
Fig. \ref{f:mirr}. Specifically, one could try to sum all diagrams that can be obtained 
from the 2nd diagram at the right-hand-side in Fig. \ref{f:mirr} by replacing the
bare response functions $G_0$ by diagrams that appear in the diagrammatic
expansion for the full response function, Eq. (\ref{Gseries}). Such replacements will
be referred to as response function-like insertions. The re-summation of the 
2nd diagram at the right-hand-side in Fig. \ref{f:mirr} with all possible
response function-like insertions 
would result in the same diagram, but with 
the bare response functions $G_0$ replaced by the full response functions $G$, see 
Fig. \ref{f:bmct1}. 

In general, such re-summations look quite promising. A possible strategy 
would be to include at least some non-mode-coupling diagrams or perhaps 
a class of non-mode-coupling diagrams, with an implicit re-summation of 
all possible response function-like insertions, and to use the resulting
expression to calculate the correction to the mode-coupling (\textit{i.e.} 
self-consistent one-loop) approximation for the irreducible memory function. 
An obvious possible pitfall is double-counting some contributions. A less
obvious pitfall is that one can quite easily generate spurious, non-physical
long-time divergences. In fact, the diagram showed in Fig. \ref{f:bmct1}
provides an example of such a divergence. The origin of the divergence is
that this diagram has unrestricted integrations over intermediate times. 
This divergence is discussed in the remainder of the present subsection.
In the next subsection, Sec. \ref{beyond1}, we show that by combining 
the diagram showed in Fig. \ref{f:bmct1} with other similar diagrams this unphysical
divergence can be avoided. 

The diagram showed in Fig. \ref{f:bmct1} leads to the following
contribution to the irreducible memory function 
(the contribution to the irreducible memory function is obtained from 
the diagram showed in Fig. \ref{f:bmct1} by using translational invariance and
taking the longitudinal component of the expression corresponding to this
diagram, see Eq. (\ref{mirr})):
\begin{eqnarray}\label{bmct1}
&& \delta M^{\mathrm{irr}}_0(k;t) = 
n^2 D_0^3 
\int_0^t dt_2 \int_0^{t_2} dt_1 \int \frac{d\mathbf{k}_1 d\mathbf{k}_2}{(2\pi)^6}
v_\mathbf{k}(\mathbf{k}_1+\mathbf{k}_2,\mathbf{k}-\mathbf{k}_1-\mathbf{k}_2) 
\\ && \times
G(|\mathbf{k}_1+\mathbf{k}_2|;t-t_2)
|\mathbf{k}_1+\mathbf{k}_2| v_{\mathbf{k}_1+\mathbf{k}_2}(\mathbf{k}_1,\mathbf{k}_2)
G(|\mathbf{k}-\mathbf{k}_1-\mathbf{k}_2|;t-t_1) S(|\mathbf{k}-\mathbf{k}_1-\mathbf{k}_2|)
\nonumber \\ && \times
G(k_2;t_2-t_1)S(k_2)  
v_{\mathbf{k}-\mathbf{k}_1}(\mathbf{k}-\mathbf{k}_1-\mathbf{k}_2,\mathbf{k}_2)
|\mathbf{k}-\mathbf{k}_1|
G(|\mathbf{k}-\mathbf{k}_1|;t_1)G(k_1;t_2) 
S(k_1) 
\nonumber \\ && \times v_\mathbf{k}(\mathbf{k}_1,\mathbf{k}-\mathbf{k}_1).
\nonumber 
\end{eqnarray}
Note that factors $|\mathbf{k}_1+\mathbf{k}_2|$ and $|\mathbf{k}-\mathbf{k}_1|$
originate from the definition of the vertices, Eqs. (\ref{V12}-\ref{V21}). 
Similar factors will appear below in Eqs. (\ref{bmct3}) and (\ref{bmct6}). 

One can show that if the full response function $G$
develops a long-lived plateau, the contribution to the irreducible memory function
given by the diagram showed in Fig. \ref{f:bmct1} grows with time rather than
exhibits a plateau. In particular, if the full response function acquires
a non-vanishing long time limit, $\lim_{t\to\infty} G(k;t) = f(k)$,
the contribution to the irreducible memory function resulting from 
the diagram showed in Fig. \ref{f:bmct1} diverges as $t^2$ as $t$ increases:
\begin{eqnarray}\label{bmct2}
&& \delta M^{\mathrm{irr}}_0(k;t) = 
\frac{1}{2} t^2 n^2 D_0^3 \int \frac{d\mathbf{k}_1 d\mathbf{k}_2}{(2\pi)^6}
v_\mathbf{k}(\mathbf{k}_1+\mathbf{k}_2,\mathbf{k}-\mathbf{k}_1-\mathbf{k}_2) 
f(|\mathbf{k}_1+\mathbf{k}_2|)
\\ && \nonumber \times
|\mathbf{k}_1+\mathbf{k}_2| v_{\mathbf{k}_1+\mathbf{k}_2}(\mathbf{k}_1,\mathbf{k}_2)
f(|\mathbf{k}-\mathbf{k}_1-\mathbf{k}_2|)
S(|\mathbf{k}-\mathbf{k}_1-\mathbf{k}_2|)f(k_2) S(k_2) 
\\ \nonumber && \times
v_{\mathbf{k}-\mathbf{k}_1}(\mathbf{k}-\mathbf{k}_1-\mathbf{k}_2,\mathbf{k}_2)
|\mathbf{k}-\mathbf{k}_1|
f(|\mathbf{k}-\mathbf{k}_1|)f(k_1) 
S(k_1) 
v_\mathbf{k}(\mathbf{k}_1,\mathbf{k}-\mathbf{k}_1) + o(t^2)
\end{eqnarray}
As mentioned above, the origin of the leading term Eq. (\ref{bmct2})
is the fact that integrations over intermediate times are unrestricted. 

The problem described above forces us to be a little more careful while
calculating corrections to the irreducible memory function.  
In the next two subsections we consider corrections originating from 
two classes of non-mode-coupling diagrams. The first class includes,
among others, diagrams that can be obtained from
the 2nd diagram at the right-hand-side of the diagrammatic equation 
in Fig. \ref{f:mirr} by replacing the
bare response functions $G_0$ by diagrams that appear in the diagrammatic
expansion for the full response function. The second class includes, among others,
diagrams that can be obtained from
the 3rd diagram at the right-hand-side of Fig. \ref{f:mirr} by replacing the
bare response functions $G_0$ by diagrams that appear in the diagrammatic
expansion for the full response function. We show that by re-summing
each of the two classes of 
diagrams we get well-behaving corrections to the irreducible memory function.

\subsection{The first correction}\label{beyond1}

If one propagator line is cut in the 2nd diagram 
in the expansion showed in Fig. \ref{f:mirr}, this diagram turns into
the 1st diagram in this expansion, \textit{i.e} into a mode-coupling diagram.
As the first correction we will re-sum the following well defined class of  
diagrams: all diagrams that turn into mode-coupling diagrams  
contributing to the irreducible memory function upon removing one response function-like 
insertion. In other words, these are the diagrams that contribute to the 
irreducible memory function and 
separate into two disconnected pieces upon removing the left and right vertices
and cutting through a single propagator line. In addition, we will impose the following
additional condition: if the response function-like insertion that makes the
diagram a non-mode-coupling diagram is removed together with its beginning 
and ending vertices, there should be no continuous path from the
the right vertex to the left vertex.

\begin{figure}
\centerline{\includegraphics[scale=.17]{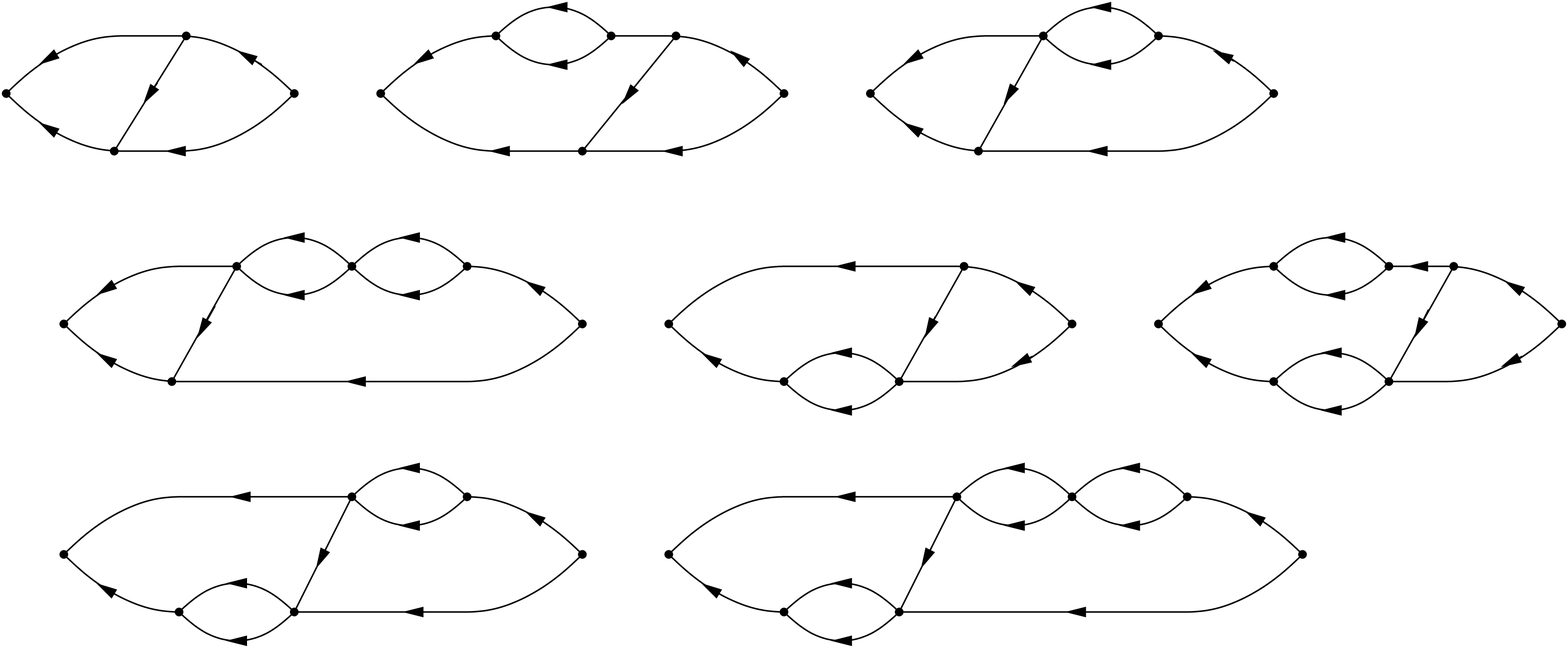}}
\caption{Example diagrams that separate into two disconnected pieces upon removing the 
left and right vertices and cutting through a single propagator line. 
These diagrams have an additional property: 
if the response function-like insertion which makes the
diagram a non-mode-coupling diagram is removed together with its beginning 
and ending vertices, there is no continuous path from the
the right vertex to the left vertex.}
\label{f:bmct2}
\end{figure}

\begin{figure}
\centerline{\includegraphics[scale=.17]{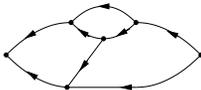}}
\caption{An example diagram that turns into a mode-coupling diagram upon removing the 
left and right vertices and cutting through a single propagator line. 
For this diagram, if the 
response function-like insertion which makes the
diagram a non-mode-coupling diagram is removed together with its beginning 
and ending vertices, there is still continuous path from the
the right vertex to the left vertex.}
\label{f:bmct3}
\end{figure}

In Fig. \ref{f:bmct2} we show a few representative diagrams that are to be re-summed.
While performing the re-summation one has to remember that the diagrams 
with odd and even number of four-leg vertices contribute with negative and
positive sign, respectively.  
In Fig. \ref{f:bmct3} we show an example of a diagram which 
separates into two disconnected pieces upon removing the left and right vertices
and cutting through a single propagator line but which does not have the
additional property described above. Diagrams similar to that in Fig. \ref{f:bmct3} 
are not included in the re-summation proposed here. The main reason for this additional
requirement is the simplicity of the resulting expressions. 
Including all non-mode-coupling diagrams 
that turn into mode-coupling diagrams upon removing one response function-like 
insertion (\textit{e.g.} including diagram showed in Fig. \ref{f:bmct3} and 
similar diagrams) is more complicated and will be discussed elsewhere \cite{GSHHEF}.

\begin{figure}
\centerline{\includegraphics[scale=.17]{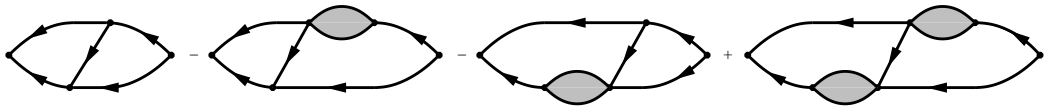}}
\caption{The result of the re-summation of diagrams that separate into 
two disconnected pieces upon removing the left and right vertices and cutting 
through a single propagator line. 
The re-summed diagrams have an additional property: 
if the response function-like insertion which makes the
diagram a non-mode-coupling diagram is removed together with its beginning 
and ending vertices, there is no continuous path from the
the right vertex to the left vertex.}
\label{f:bmct4}
\end{figure}

The result of the re-summation of the above described class of 
diagrams is showed in Fig. \ref{f:bmct4} (bubble insertions in this figure are the
memory function matrices defined in Eq. (\ref{Mredseries}) 
and illustrated in Fig. \ref{f:mred}). Briefly, the first two diagrams 
showed in Fig. \ref{f:bmct2} contribute to the first diagram in Fig. \ref{f:bmct4}.
This diagram is identical to the diagram showed in Fig. \ref{f:bmct1}. 
The third and fourth diagrams in Fig. \ref{f:bmct2} contribute to the second
diagram in Fig. \ref{f:bmct4}. The fifth and sixth 
diagrams in Fig. \ref{f:bmct2} contribute to the third
diagram in Fig. \ref{f:bmct4}. Finally, the seventh and eighth 
diagrams in Fig. \ref{f:bmct2} contribute to the fourth
diagram in Fig. \ref{f:bmct4}. 

The four diagrams showed in Fig. \ref{f:bmct4} lead to the following contribution
to the irreducible memory function:
\begin{eqnarray}\label{bmct3}
&& \delta M^{\mathrm{irr}}_1(k;t) = 
n^2 D_0^3 
\int_0^t dt_4 \int_0^{t_4} dt_3 \int_0^{t_3} dt_2  \int_0^{t_2} dt_1
\int \frac{d\mathbf{k}_1 d\mathbf{k}_2}{(2\pi)^6}
v_\mathbf{k}(\mathbf{k}_1+\mathbf{k}_2,\mathbf{k}-\mathbf{k}_1-\mathbf{k}_2) 
\nonumber \\ && \times
G(|\mathbf{k}_1+\mathbf{k}_2|;t-t_4)
(\delta(t_4-t_3)-M(|\mathbf{k}_1+\mathbf{k}_2|;t_4-t_3))
|\mathbf{k}_1+\mathbf{k}_2| v_{\mathbf{k}_1+\mathbf{k}_2}(\mathbf{k}_1,\mathbf{k}_2)
\nonumber \\ && \times
G(|\mathbf{k}-\mathbf{k}_1-\mathbf{k}_2|;t-t_1)
S(|\mathbf{k}-\mathbf{k}_1-\mathbf{k}_2|)G(k_2;t_3-t_2) S(k_2)  
\nonumber \\ && \times
v_{\mathbf{k}-\mathbf{k}_1}(\mathbf{k}-\mathbf{k}_1-\mathbf{k}_2,\mathbf{k}_2)
|\mathbf{k}-\mathbf{k}_1|
(\delta(t_2-t_1)-M(|\mathbf{k}-\mathbf{k}_1|;t_2-t_1)) 
G(|\mathbf{k}-\mathbf{k}_1|;t_1)  
\nonumber \\ && \times 
G(k_1;t_3)  S(k_1)
v_\mathbf{k}(\mathbf{k}_1,\mathbf{k}-\mathbf{k}_1).
\end{eqnarray}
To show that these four diagrams give a well-behaving contribution we will first
rewrite Eq. (\ref{bmct3}). To this end we will use the
following two identities which can be obtained from Eq. (\ref{Fm}),
\begin{eqnarray}\label{der}
&& \int_{0}^t dt_1 \left(\delta(t-t_1) - M(k;t-t_1)\right)
G(k;t_1) 
\\ && \nonumber = \int_{0}^t dt_1 G(k;t-t_1) \left(\delta(t_1) - M(k;t_1)\right)
= -\frac{S(k)}{D_0 k^2}  \partial_t G(k;t).
\end{eqnarray}
These identities allow us to rewrite Eq. (\ref{bmct3}) in the following form
\begin{eqnarray}\label{bmct4}
&& \delta M^{\mathrm{irr}}_1(k;t) = 
n^2 D_0 \int_0^{t} dt_3 \int_0^{t_3} dt_2 
\int \frac{d\mathbf{k}_1 d\mathbf{k}_2}{(2\pi)^6}
v_\mathbf{k}(\mathbf{k}_1+\mathbf{k}_2,\mathbf{k}-\mathbf{k}_1-\mathbf{k}_2) 
\nonumber \\ && \times
\partial_{t} G(|\mathbf{k}_1+\mathbf{k}_2|;t-t_3)
S(|\mathbf{k}_1+\mathbf{k}_2|)
\tilde{v}_{\mathbf{k}_1+\mathbf{k}_2}(\mathbf{k}_1,\mathbf{k}_2)
G(|\mathbf{k}-\mathbf{k}_1-\mathbf{k}_2|;t-t_1)
\nonumber \\ && \times 
S(|\mathbf{k}-\mathbf{k}_1-\mathbf{k}_2|)
G(k_2;t_3-t_2)S(k_2)  
\tilde{v}_{\mathbf{k}-\mathbf{k}_1}(\mathbf{k}-\mathbf{k}_1-\mathbf{k}_2,\mathbf{k}_2)
\nonumber \\ && \times \partial_{t_2} 
G(|\mathbf{k}-\mathbf{k}_1|;t_2)
S(|\mathbf{k}-\mathbf{k}_1|)
G(k_1;t_3) S(k_1) v_\mathbf{k}(\mathbf{k}_1,\mathbf{k}-\mathbf{k}_1),
\end{eqnarray}
where modified vertex function $\tilde{v}$ is defined in Eq. (\ref{vdef1}).

Now we can appreciate the effect of adding to the first diagram in Fig. \ref{f:bmct4}
(which is identical to the diagram  showed in Fig. \ref{f:bmct1} and discussed in the 
previous subsection) the remaining three diagrams. Roughly speaking, by adding the 
additional diagrams two response functions in the first diagram in 
Fig. \ref{f:bmct4} get replaced by time derivatives of response functions. 
As a result, restrictions for integrations over intermediate times are introduced. 
To see this we need to recognize
the fact that even if the full response function develops a long-lived plateau,
its time derivative still decays fast. In particular, if the full response
function acquires a non-vanishing long-time limit, its time derivative can be trivially
integrated over time. Thus, if $\lim_{t\to\infty} G(k;t) = f(k)$, then the
long-time limit of correction (\ref{bmct4}) is finite and given by the 
following expression:
\begin{eqnarray}\label{bmct5}
\nonumber 
&& \lim_{t\to\infty} \delta M^{\mathrm{irr}}_1(k;t) = 
n^2 D_0 \int_0^{t} dt_3 \int_0^{t_3} dt_2 
\int \frac{d\mathbf{k}_1 d\mathbf{k}_2}{(2\pi)^6}
v_\mathbf{k}(\mathbf{k}_1+\mathbf{k}_2,\mathbf{k}-\mathbf{k}_1-\mathbf{k}_2) 
\nonumber \\ && \times
(1-f(|\mathbf{k}_1+\mathbf{k}_2|))
S(|\mathbf{k}_1+\mathbf{k}_2|)
\tilde{v}_{\mathbf{k}_1+\mathbf{k}_2}(\mathbf{k}_1,\mathbf{k}_2)
f(|\mathbf{k}-\mathbf{k}_1-\mathbf{k}_2|)
\nonumber \\ && \times 
S(|\mathbf{k}-\mathbf{k}_1-\mathbf{k}_2|)
f(k_2)S(k_2)  
\tilde{v}_{\mathbf{k}-\mathbf{k}_1}(\mathbf{k}-\mathbf{k}_1-\mathbf{k}_2,\mathbf{k}_2)
\nonumber \\ && \times  
(1-f(|\mathbf{k}-\mathbf{k}_1|)
S(|\mathbf{k}-\mathbf{k}_1|)
f(k_1) S(k_1) v_\mathbf{k}(\mathbf{k}_1,\mathbf{k}-\mathbf{k}_1).
\end{eqnarray}

It is instructive to derive from the above expression the contribution
to the function $m$ defined in Eq. (\ref{mdef}):
\begin{eqnarray}\label{bmct5a}
\nonumber 
&& \delta m_1(k) = \lim_{t\to\infty} \frac{S(k)}{D_0 k^2} \delta M^{\mathrm{irr}}_1(k;t)
= n^2 S(k) \int \frac{d\mathbf{k}_1 d\mathbf{k}_2}{(2\pi)^6}
\tilde{v}_\mathbf{k}(\mathbf{k}_1+\mathbf{k}_2,\mathbf{k}-\mathbf{k}_1-\mathbf{k}_2) 
\nonumber \\ && \times
(1-f(|\mathbf{k}_1+\mathbf{k}_2|))
S(|\mathbf{k}_1+\mathbf{k}_2|)
\tilde{v}_{\mathbf{k}_1+\mathbf{k}_2}(\mathbf{k}_1,\mathbf{k}_2)
f(|\mathbf{k}-\mathbf{k}_1-\mathbf{k}_2|)
\nonumber \\ && \times 
S(|\mathbf{k}-\mathbf{k}_1-\mathbf{k}_2|)
f(k_2)S(k_2)  
\tilde{v}_{\mathbf{k}-\mathbf{k}_1}(\mathbf{k}-\mathbf{k}_1-\mathbf{k}_2,\mathbf{k}_2)
\nonumber \\ && \times  
(1-f(|\mathbf{k}-\mathbf{k}_1|)
S(|\mathbf{k}-\mathbf{k}_1|)
f(k_1) S(k_1) \tilde{v}_\mathbf{k}(\mathbf{k}_1,\mathbf{k}-\mathbf{k}_1).
\end{eqnarray}
It can be seen that the above expression can be interpreted as a renormalized
diagram. The vertices of this diagram are given by the modified
vertex functions $\tilde{v}_{\mathbf{k}}$ and the bonds are equal to either $f(k)S(k)$ or
$(1-f(k))S(k)$. Alternatively, using relation (\ref{mctnep}) between 
$f(k)$ and $m(k)$, the above expression can be re-written in such a way
that the internal vertices of the renormalized diagram are given by 
$\tilde{v}_{\mathbf{k}}/m(k)$ and all the bonds are equal to $f(k)S(k)$. 
We will see in the next 
section that the expression for the second correction can be written in a similar way.

\textit{A priori}, it is not clear whether the above expression is positive or
negative, \textit{i.e.} whether it moves the ergodicity breaking transition of
the standard mode-coupling theory 
towards higher or lower temperatures (or lower or higher volume fractions),
respectively. The explicit calculation described in subsection \ref{beyond3}
suggests that expression (\ref{bmct5}) gives a small, positive contribution 
to the irreducible memory function. 

\subsection{The second correction}\label{beyond2}

If one propagator line is cut in 
the 3rd diagram in the expansion showed in Fig. \ref{f:mirr}, this diagram 
turns into one of the mode-coupling diagrams contributing to the
\emph{reducible} memory matrix. Specifically, by removing one propagator 
line we can turn the 3rd diagram in the expansion in Fig. \ref{f:mirr}
into the second diagram in the expansion in Fig. \ref{f:mred}.  
As the second correction we will re-sum the 
following class of diagrams: all diagrams that turn into mode-coupling
diagrams contributing to the memory matrix upon removing one response 
function-like insertion. In addition, we will impose an additional condition
which is similar to that introduce in Sec. \ref{beyond1}. In the present case
the description of this additional condition is a little more complicated,
but the idea is the same. We will re-sum only those diagrams for which,
if the response function-like insertion which makes the
diagram a non-mode-coupling diagram is removed together with its 
beginning and ending vertices, there is only one continuous path 
from the right vertex to the closest articulation four-leg
vertex and only one continuous path from the left vertex to the 
closest articulation four-leg vertex.

\begin{figure}
\centerline{\includegraphics[scale=.17]{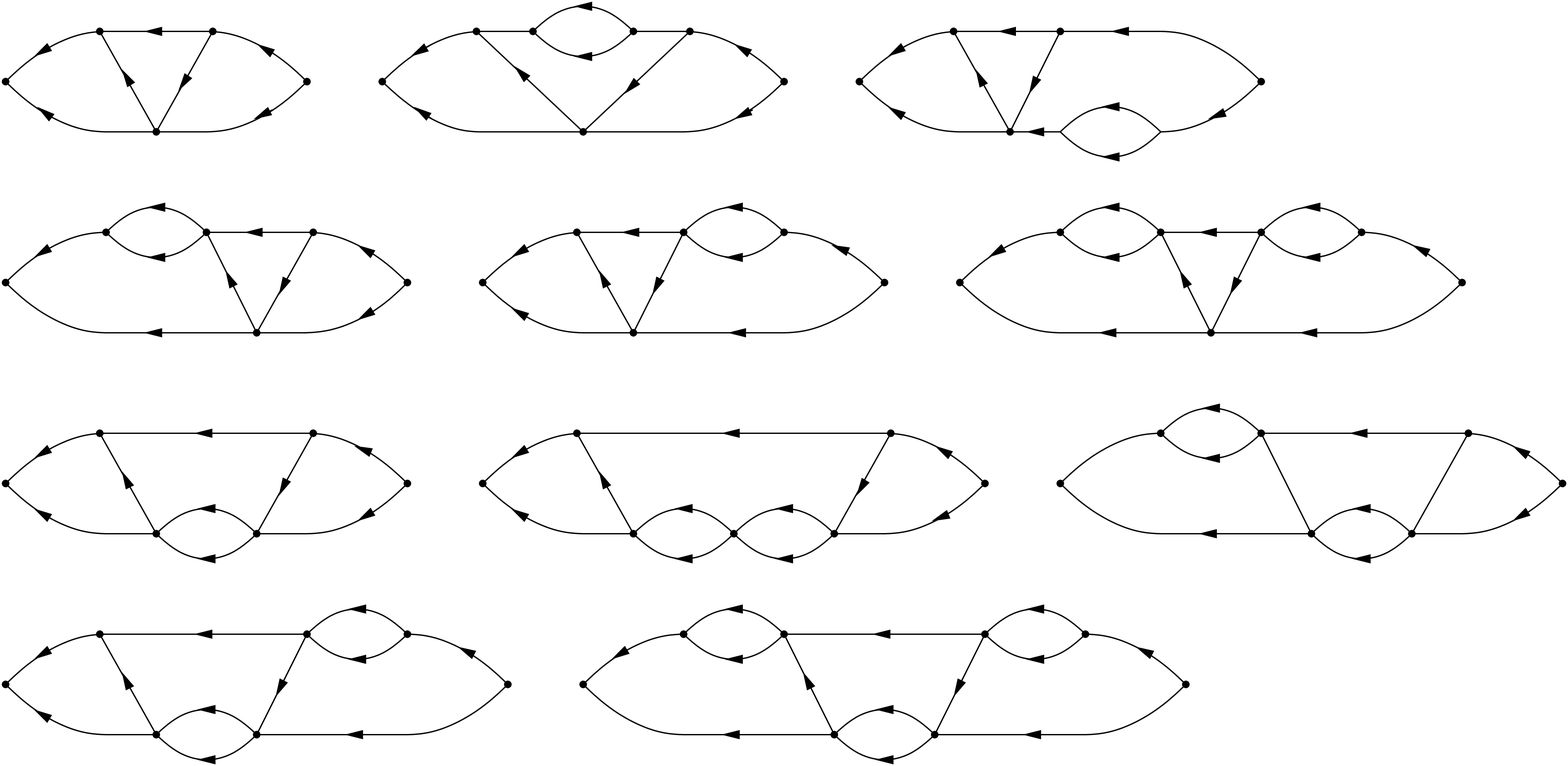}}
\caption{Example diagrams that turn into mode-coupling contributions to
the memory matrix upon removing one response function-like insertion. 
These diagrams have an additional property: 
if the response function-like insertion which makes the
diagram a non-mode-coupling diagram is removed together with its 
beginning and ending vertices, there is only one continuous path 
from the right vertex to the closest articulation four-leg
vertex and only one continuous path from the left vertex to the 
closest articulation four-leg vertex.}
\label{f:bmct5}
\end{figure}

\begin{figure}
\centerline{\includegraphics[scale=.17]{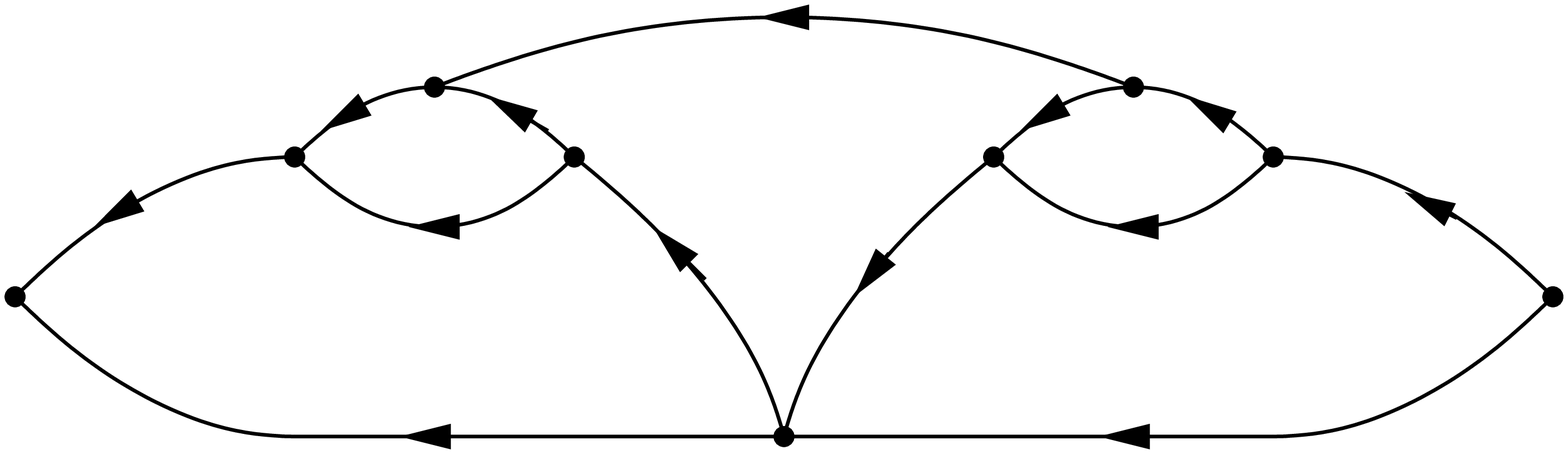}}
\caption{An example diagram that turns into mode-coupling-like contributions to
the memory matrix upon removing one response function-like 
insertion. For this diagram, 
if the response function-like insertion which makes the
diagram a non-mode-coupling diagram is removed together with its 
beginning and ending vertices, there are two continuous paths 
from the right vertex to the closest articulation four-leg
vertex (which is the only four-leg vertex in this diagram) 
and two continuous paths from the left vertex to the 
closest articulation four-leg vertex.}
\label{f:bmct6}
\end{figure}

In Fig. \ref{f:bmct5} we show a few representative diagrams that are to be re-summed.
Again, while performing the re-summation one has to remember that the diagrams 
with odd and even number of four-leg vertices contribute with negative and
positive sign, respectively.  
In Fig. \ref{f:bmct6} we show an example of a diagram which does not 
have the additional property described in the preceding paragraph. In the 
diagram showed in Fig. \ref{f:bmct6}, 
if the response function-like insertion together with its 
beginning and ending vertices is removed, there are two continuous paths 
from the right vertex to the closest articulation four-leg
vertex (which is the only four-leg vertex in this diagram) 
and two continuous paths from the left vertex to the 
closest articulation four-leg vertex.
Diagrams similar to that in Fig. \ref{f:bmct6} 
are not included in the re-summation proposed here. Again, 
the main reason for this additional
requirement is the simplicity of the resulting expressions. 

\begin{figure}
\centerline{\includegraphics[scale=.17]{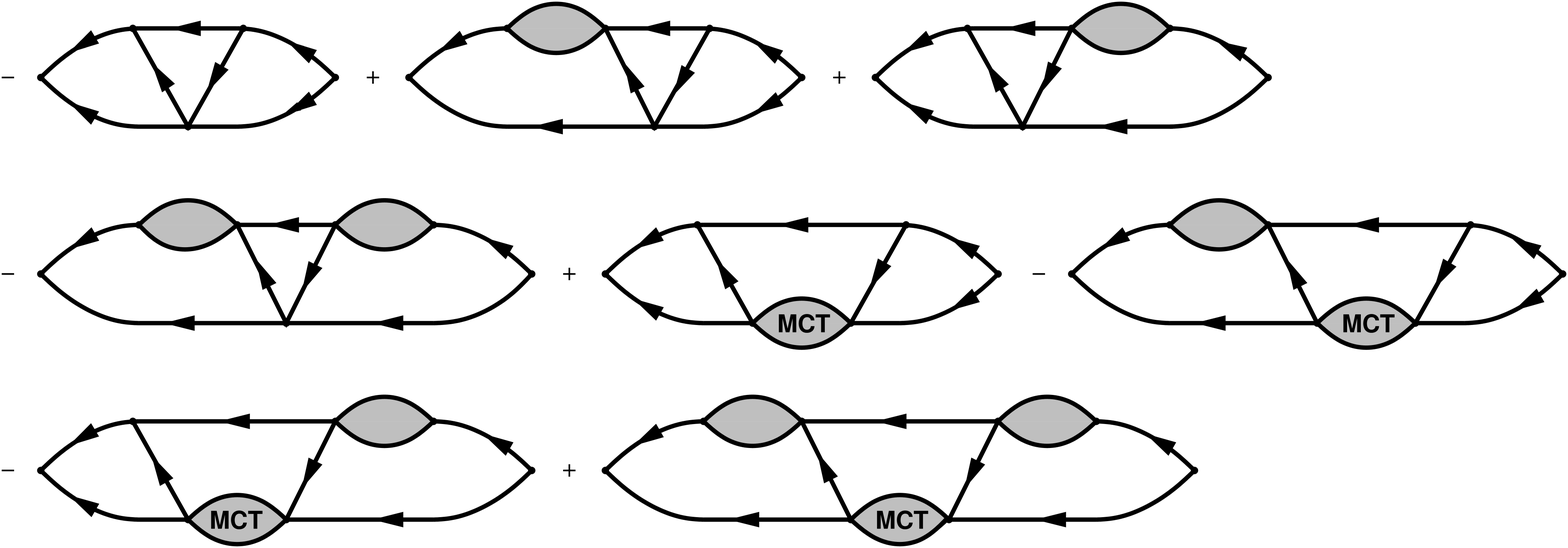}}
\caption{The result of the re-summation of diagrams that 
turn into mode-coupling-like contributions to
the memory function upon removing one response function-like 
insertion. 
The re-summed diagrams have an additional property: 
if the response function-like insertion together with its 
beginning and ending vertices is removed, there is only one continuous path 
from the right vertex to the closest articulation four-leg
vertex and only one continuous path from the left vertex to the 
closest articulation four-leg vertex.}
\label{f:bmct7}
\end{figure}

\begin{figure}
\centerline{\includegraphics[scale=.17]{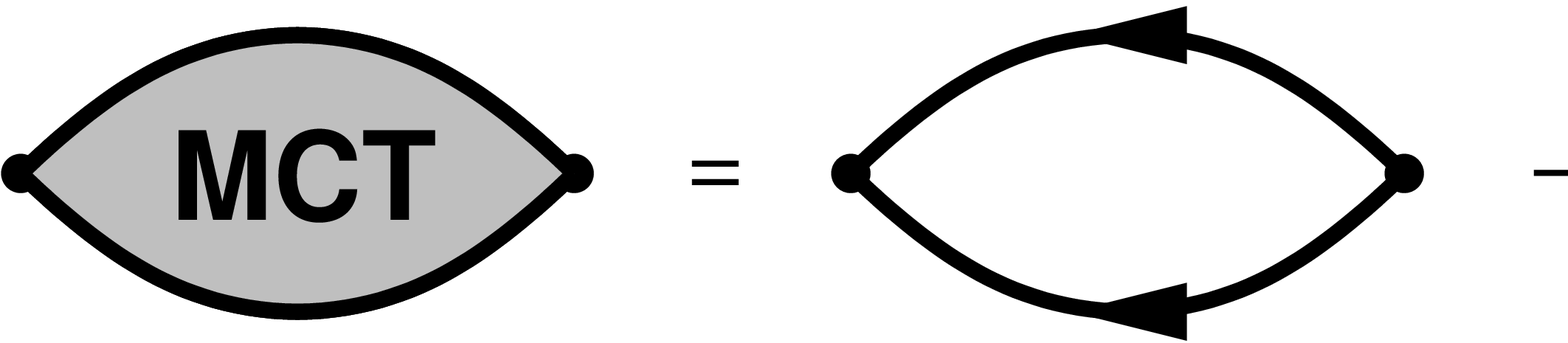}}
\caption{Memory matrix $\mathbf{M}$ calculated within the mode-coupling approximation.}
\label{f:bmct8}
\end{figure}

The result of the re-summation of the above described class of 
diagrams is showed in Fig. \ref{f:bmct7}. Again, unlabeled bubble insertions 
are the memory function matrices defined in Eq. (\ref{Mredseries}) 
and illustrated in Fig. \ref{f:mred}. In contrast, bubble insertions labeled
MCT are the memory function matrices within the mode-coupling approximation showed
in Fig. \ref{f:bmct8}. The presence of the 
latter insertions are the consequence of the definition
of the class of diagrams that are re-summed in this subsection. Specifically,
we imposed the requirement 
that after one response function-like insertion is removed, the
resulting diagram was a mode-coupling diagram contributing to the memory matrix.
The last condition means that after the response function-like insertion is
removed (but its beginning and ending vertices are kept) the resulting diagram
has to have the following property: if the left and right vertices, and the four-leg
articulation vertices are removed from the diagram, each part that
used to be between successive articulation vertices has to consist of 
two disconnected pieces. 

Briefly, the first three diagrams 
showed in Fig. \ref{f:bmct5} contribute to the first diagram in Fig. \ref{f:bmct7}.
The fourth, fifth and sixth diagrams in Fig. \ref{f:bmct5} contribute to the second,
third, and fourth diagrams in Fig. \ref{f:bmct7}, respectively. 
The seventh and eighth 
diagrams in Fig. \ref{f:bmct5} contribute to the fifth
diagram in Fig. \ref{f:bmct7}. Finally, the
ninth, tenth and eleventh diagrams in Fig. \ref{f:bmct5} contribute to the sixth,
seventh, and eighth diagrams in Fig. \ref{f:bmct7}, respectively. 

The eight diagrams showed in Fig. \ref{f:bmct7} lead to the following contribution
to the irreducible memory function:
\begin{eqnarray}\label{bmct6}
&& \delta M^{\mathrm{irr}}_2(k;t) = 
- n^3 D_0^4 
\int_0^t dt_6 \int_0^{t_6} dt_5 
\int_0^{t_5} dt_4 \int_0^{t_4} dt_3 \int_0^{t_3} dt_2  \int_0^{t_2} dt_1
\int \frac{d\mathbf{k}_1 d\mathbf{k}_2 d\mathbf{k}_3}{(2\pi)^9}
\nonumber \\ && \times
v_\mathbf{k}(\mathbf{k}_1+\mathbf{k}_2-\mathbf{k}_3,
\mathbf{k}-\mathbf{k}_1-\mathbf{k}_2+\mathbf{k}_3) 
G(|\mathbf{k}_1+\mathbf{k}_2-\mathbf{k}_3|;t-t_4)
S(|\mathbf{k}_1+\mathbf{k}_2-\mathbf{k}_3|) 
\nonumber \\ && \times
G(|\mathbf{k}-\mathbf{k}_1-\mathbf{k}_2+\mathbf{k}_3|;t-t_6)
(\delta(t_6-t_5)-M(|\mathbf{k}-\mathbf{k}_1-\mathbf{k}_2+\mathbf{k}_3|;t_6-t_5))
\nonumber \\ &&
|\mathbf{k}-\mathbf{k}_1-\mathbf{k}_2+\mathbf{k}_3|
v_{\mathbf{k}-\mathbf{k}_1-\mathbf{k}_2+\mathbf{k}_3}
(\mathbf{k}-\mathbf{k}_1-\mathbf{k}_2,\mathbf{k}_3)G(k_3;t_5-t_4)S(k_3)
\nonumber \\ && 
G(|\mathbf{k}-\mathbf{k}_1-\mathbf{k}_2|;t_5-t_2)
S(|\mathbf{k}-\mathbf{k}_1-\mathbf{k}_2|)
\nonumber \\ && 
\mathbf{v}(\mathbf{k}_1+\mathbf{k}_2-\mathbf{k}_3,\mathbf{k}_3)\cdot
\left[\mathcal{I}\delta(t_4-t_3) - 
\mathbf{M}_{\mathrm{MCT}}(\mathbf{k}_1+\mathbf{k}_2;t_4-t_3)\right]\cdot
\mathbf{v}(\mathbf{k}_1,\mathbf{k}_2)
\nonumber \\ && 
G(k_2;t_3-t_2)S(k_2)
v_{\mathbf{k}-\mathbf{k}_1}(\mathbf{k}_2,\mathbf{k}-\mathbf{k}_1-\mathbf{k}_2)
|\mathbf{k}-\mathbf{k}_1|
\nonumber \\ && 
(\delta(t_2-t_1)-M(|\mathbf{k}-\mathbf{k}_1|;t_2-t_1))
G(|\mathbf{k}-\mathbf{k}_1|;t_1)G(k_1;t_3) S(k_1) 
v_\mathbf{k}(\mathbf{k}_1,\mathbf{k}-\mathbf{k}_1),
\nonumber \\ 
\end{eqnarray}
where $\mathcal{I}$ denotes the unit tensor.
To write down Eq. (\ref{bmct6}) in a slightly more compact form 
we used the function $\mathbf{v}$ 
defined in Eq. (\ref{vbdef}) and we introduced the
mode-coupling theory's memory matrix that has the delta function originating
from translational invariance factored out,
\begin{eqnarray}\label{bmct6a}
\mathbf{M}_{\mathrm{MCT}}(\mathbf{k},\mathbf{k}_1;t) = 
\mathbf{M}_{\mathrm{MCT}}(\mathbf{k};t)
(2\pi)^3 \delta(\mathbf{k}-\mathbf{k}_1),
\end{eqnarray}
where $\mathbf{M}_{\mathrm{MCT}}(\mathbf{k},\mathbf{k}_1;t)$ 
is the memory matrix calculated within the mode-coupling approximation 
(see Fig. \ref{f:bmct8}). For future use (see Eq. (\ref{bmct8a}) below) 
we also define mode-coupling theory's 
irreducible memory matrix with the delta function part factored out,
\begin{eqnarray}\label{bmct6b}
\mathbf{M}_{\mathrm{MCT}}^{\mathrm{irr}}(\mathbf{k},\mathbf{k}_1;t) = 
\mathbf{M}_{\mathrm{MCT}}^{\mathrm{irr}}(\mathbf{k};t)
(2\pi)^3 \delta(\mathbf{k}-\mathbf{k}_1).
\end{eqnarray}
It should be noted that we use the same symbols for memory matrices
with and without delta function factors. Whenever we use memory matrices
with delta functions factored out, we will indicate this fact by specifying 
their arguments.

As in the previous subsection, we can use identities (\ref{der}) to 
replace the memory functions (but not the mode-coupling memory matrix)
by time derivatives,
\begin{eqnarray}\label{bmct7}
&& \delta M^{\mathrm{irr}}_2(k;t) = 
- n^3 D_0^2 
\int_0^5 dt_5 
\int_0^{t_5} dt_4 \int_0^{t_4} dt_3 \int_0^{t_3} dt_2 
\int \frac{d\mathbf{k}_1 d\mathbf{k}_2 d\mathbf{k}_3}{(2\pi)^9}
\nonumber \\ && \times
v_\mathbf{k}(\mathbf{k}_1+\mathbf{k}_2-\mathbf{k}_3,
\mathbf{k}-\mathbf{k}_1-\mathbf{k}_2+\mathbf{k}_3) 
G(|\mathbf{k}_1+\mathbf{k}_2-\mathbf{k}_3|;t-t_4)
S(|\mathbf{k}_1+\mathbf{k}_2-\mathbf{k}_3|) 
\nonumber \\ && \times
\partial_t G(|\mathbf{k}-\mathbf{k}_1-\mathbf{k}_2+\mathbf{k}_3|;t-t_5)
S(|\mathbf{k}-\mathbf{k}_1-\mathbf{k}_2+\mathbf{k}_3|)
\nonumber \\ && \times
|\mathbf{k}-\mathbf{k}_1-\mathbf{k}_2+\mathbf{k}_3|
\tilde{v}_{\mathbf{k}-\mathbf{k}_1-\mathbf{k}_2+\mathbf{k}_3}
(\mathbf{k}-\mathbf{k}_1-\mathbf{k}_2,\mathbf{k}_3)G(k_3;t_5-t_4)S(k_3)
\nonumber \\ && \times
G(|\mathbf{k}-\mathbf{k}_1-\mathbf{k}_2|;t_5-t_2)
S(|\mathbf{k}-\mathbf{k}_1-\mathbf{k}_2|)
\nonumber \\ && \times
\mathbf{v}(\mathbf{k}_1+\mathbf{k}_2-\mathbf{k}_3,\mathbf{k}_3)\cdot
\left[\mathcal{I}\delta(t_4-t_3) - 
\mathbf{M}_{\mathrm{MCT}}(\mathbf{k}_1+\mathbf{k}_2;t_4-t_3)\right]\cdot
\mathbf{v}(\mathbf{k}_1,\mathbf{k}_2)
\nonumber \\ && \times
G(k_2;t_3-t_2)S(k_2)
\tilde{v}_{\mathbf{k}-\mathbf{k}_1}(\mathbf{k}_2,\mathbf{k}-\mathbf{k}_1-\mathbf{k}_2)
\nonumber \\ && \times
\partial_{t_2} 
G(|\mathbf{k}-\mathbf{k}_1|;t_2)S(|\mathbf{k}-\mathbf{k}_1|)G(k_1;t_3) S(k_1) 
v_\mathbf{k}(\mathbf{k}_1,\mathbf{k}-\mathbf{k}_1),
\end{eqnarray}
where the modified vertex function $\tilde{v}$ is defined in Eq. (\ref{vdef1}).

Again, the above expression has a well defined, finite long-time limit, even if
the full response function does not decay. To see this we need to 
recognize two facts. First, as before, 
the presence of time derivatives introduces restrictions
for integrations over intermediate times. Second, 
the term $\left[\mathcal{I}\delta(t) - 
\mathbf{M}_{\mathrm{MCT}}(\mathbf{k};t)\right]$ introduces an additional 
restriction for integration over intermediate times. 
Technically, the last statement follows from the fact that if 
$\lim_{t\to\infty} G(k;t) = f(k)$, then the irreducible memory matrix 
does not decay, 
$\lim_{t\to\infty} \mathbf{M}_{\mathrm{MCT}}^{\mathrm{irr}}(\mathbf{k};t)\neq 0$, 
and consequently the Laplace transform of this term vanishes as $z\to 0$,
\begin{eqnarray}\label{bmct8a}
\mathcal{I} - \mathbf{M}_{\mathrm{MCT}}(\mathbf{k};z) = 
\left[\mathcal{I} + \mathbf{M}_{\mathrm{MCT}}^{\mathrm{irr}}(\mathbf{k};z) \right]^{-1} 
\to 
z \left[ \lim_{t\to\infty}
\mathbf{M}_{\mathrm{MCT}}^{\mathrm{irr}}(\mathbf{k};t)\right]^{-1}
+ o(z).
\nonumber \\
\end{eqnarray}
The presence of both restrictions on integrations over intermediate times 
makes the long-time limit of correction (\ref{bmct7}) well defined. We can
show that if $\lim_{t\to\infty} G(k;t) = f(k)$ then the 
long-time limit of correction (\ref{bmct7}) is given by the 
following expression: 
\begin{eqnarray}\label{bmct8}
&& \delta M^{\mathrm{irr}}_2(k;t) = 
- n^3 D_0 
\int \frac{d\mathbf{k}_1 d\mathbf{k}_2 d\mathbf{k}_3}{(2\pi)^9}
\nonumber \\ && \times
v_\mathbf{k}(\mathbf{k}_1+\mathbf{k}_2-\mathbf{k}_3,
\mathbf{k}-\mathbf{k}_1-\mathbf{k}_2+\mathbf{k}_3) 
f(|\mathbf{k}_1+\mathbf{k}_2-\mathbf{k}_3|)
S(|\mathbf{k}_1+\mathbf{k}_2-\mathbf{k}_3|) 
\nonumber \\ && \times
(1-f(|\mathbf{k}-\mathbf{k}_1-\mathbf{k}_2+\mathbf{k}_3|))
S(|\mathbf{k}-\mathbf{k}_1-\mathbf{k}_2+\mathbf{k}_3|)
\nonumber \\ && \times
|\mathbf{k}-\mathbf{k}_1-\mathbf{k}_2+\mathbf{k}_3|
\tilde{v}_{\mathbf{k}-\mathbf{k}_1-\mathbf{k}_2+\mathbf{k}_3}
(\mathbf{k}-\mathbf{k}_1-\mathbf{k}_2,\mathbf{k}_3)f(k_3)S(k_3)
\nonumber \\ &&  \times
f(|\mathbf{k}-\mathbf{k}_1-\mathbf{k}_2|)
S(|\mathbf{k}-\mathbf{k}_1-\mathbf{k}_2|)
S(|\mathbf{k}_1+\mathbf{k}_2|)
\nonumber \\ &&  \times
\left[ \tilde{v}_{\mathbf{k}_1+\mathbf{k}_2}
(\mathbf{k}_1+\mathbf{k}_2-\mathbf{k}_3,\mathbf{k}_3)
\tilde{v}_{\mathbf{k}_1+\mathbf{k}_2}(\mathbf{k}_1,\mathbf{k}_2)
m^{-1}_{\mathrm{MCT}}(|\mathbf{k}_1+\mathbf{k}_2|)
\right. \nonumber \\ && \left. 
+ \left(\tilde{\mathbf{v}}(\mathbf{k}_1+\mathbf{k}_2-\mathbf{k}_3,\mathbf{k}_3)\cdot
\tilde{\mathbf{v}}(\mathbf{k}_1,\mathbf{k}_2)-
\tilde{v}_{\mathbf{k}_1+\mathbf{k}_2}
(\mathbf{k}_1+\mathbf{k}_2-\mathbf{k}_3,\mathbf{k}_3)
\tilde{v}_{\mathbf{k}_1+\mathbf{k}_2}(\mathbf{k}_1,\mathbf{k}_2)\right) 
\right. \nonumber \\ && \left.
\times m^{-1}_{\mathrm{tMCT}}(|\mathbf{k}_1+\mathbf{k}_2|) \right]
f(k_2)S(k_2)
\tilde{v}_{\mathbf{k}-\mathbf{k}_1}(\mathbf{k}_2,\mathbf{k}-\mathbf{k}_1-\mathbf{k}_2)
\nonumber \\ &&  \times
(1-f(|\mathbf{k}-\mathbf{k}_1|))S(|\mathbf{k}-\mathbf{k}_1|)f(k_1) S(k_1) 
v_\mathbf{k}(\mathbf{k}_1,\mathbf{k}-\mathbf{k}_1).
\end{eqnarray}
Again, to write Eq. (\ref{bmct8}) in a slightly more compact form 
we used modified vertex functions $\tilde{v}$
and  $\tilde{\mathbf{v}}$ defined in Eqs. (\ref{vdef1}-\ref{vbdef1}) and function 
$m_{\mathrm{MCT}}$ defined in Eq. (\ref{mmctdef}).
Furthermore, the function $m_{\mathrm{tMCT}}$ in Eq. (\ref{bmct8}) 
is related to the transverse part of the mode-coupling theory's 
irreducible memory matrix through the following equations,
\begin{eqnarray}\label{mtmctdef0}
\left(\mathcal{I}-\hat{\mathbf{k}}\hat{\mathbf{k}}\right):
\mathbf{M}^{\mathrm{irr}}_{\mathrm{MCT}}(\mathbf{k},\mathbf{k}_1;t)
=2 M_{\mathrm{tMCT}}(k;t)(2\pi)^3\delta(\mathbf{k}-\mathbf{k}_1),
\end{eqnarray}
\begin{eqnarray}\label{mtmctdef}
m_{\mathrm{tMCT}}(k)  = 
\lim_{t\to\infty} \frac{S(k)}{D_0 k^2} M_{\mathrm{tMCT}}(k;t).
\end{eqnarray}

We shall point out that, to the best of our knowledge, the transverse part
of the irreducible memory function  has never appeared before in any theory
of the dynamics of Brownian systems. It is not entirely clear whether its
appearance in Eq. (\ref{bmct8}) is a result of one of approximations involved in 
deriving this equation or whether it has a more fundamental origin.  

As in the previous subsection, it is 
instructive to derive from the above expression the contribution
to the function $m$ defined in Eq. (\ref{mdef}):
\begin{eqnarray}\label{bmct8b}
\nonumber 
&& \delta m_2(k) = \lim_{t\to\infty} \frac{S(k)}{D_0 k^2} \delta M^{\mathrm{irr}}_2(k;t)
= - n^3 S(k) 
\int \frac{d\mathbf{k}_1 d\mathbf{k}_2 d\mathbf{k}_3}{(2\pi)^9}
\nonumber \\ && \times
\tilde{v}_\mathbf{k}(\mathbf{k}_1+\mathbf{k}_2-\mathbf{k}_3,
\mathbf{k}-\mathbf{k}_1-\mathbf{k}_2+\mathbf{k}_3) 
f(|\mathbf{k}_1+\mathbf{k}_2-\mathbf{k}_3|)
S(|\mathbf{k}_1+\mathbf{k}_2-\mathbf{k}_3|) 
\nonumber \\ && \times
(1-f(|\mathbf{k}-\mathbf{k}_1-\mathbf{k}_2+\mathbf{k}_3|))
S(|\mathbf{k}-\mathbf{k}_1-\mathbf{k}_2+\mathbf{k}_3|)
\nonumber \\ && \times
|\mathbf{k}-\mathbf{k}_1-\mathbf{k}_2+\mathbf{k}_3|
\tilde{v}_{\mathbf{k}-\mathbf{k}_1-\mathbf{k}_2+\mathbf{k}_3}
(\mathbf{k}-\mathbf{k}_1-\mathbf{k}_2,\mathbf{k}_3)f(k_3)S(k_3)
\nonumber \\ &&  \times
f(|\mathbf{k}-\mathbf{k}_1-\mathbf{k}_2|)
S(|\mathbf{k}-\mathbf{k}_1-\mathbf{k}_2|)
S(|\mathbf{k}_1+\mathbf{k}_2|)
\nonumber \\ && 
\left[ \tilde{v}_{\mathbf{k}_1+\mathbf{k}_2}
(\mathbf{k}_1+\mathbf{k}_2-\mathbf{k}_3,\mathbf{k}_3)
\tilde{v}_{\mathbf{k}_1+\mathbf{k}_2}(\mathbf{k}_1,\mathbf{k}_2)
m^{-1}_{\mathrm{MCT}}(|\mathbf{k}_1+\mathbf{k}_2|)
\right. \nonumber \\ && \left. 
+ \left(\tilde{\mathbf{v}}(\mathbf{k}_1+\mathbf{k}_2-\mathbf{k}_3,\mathbf{k}_3)\cdot
\tilde{\mathbf{v}}(\mathbf{k}_1,\mathbf{k}_2)-
\tilde{v}_{\mathbf{k}_1+\mathbf{k}_2}
(\mathbf{k}_1+\mathbf{k}_2-\mathbf{k}_3,\mathbf{k}_3)
\tilde{v}_{\mathbf{k}_1+\mathbf{k}_2}(\mathbf{k}_1,\mathbf{k}_2)\right) 
\right. \nonumber \\ && \left.
\times m^{-1}_{\mathrm{tMCT}}(|\mathbf{k}_1+\mathbf{k}_2|) \right]
f(k_2)S(k_2)
\tilde{v}_{\mathbf{k}-\mathbf{k}_1}(\mathbf{k}_2,\mathbf{k}-\mathbf{k}_1-\mathbf{k}_2)
\nonumber \\ &&  \times
(1-f(|\mathbf{k}-\mathbf{k}_1|))S(|\mathbf{k}-\mathbf{k}_1|)f(k_1) S(k_1) 
\tilde{v}_\mathbf{k}(\mathbf{k}_1,\mathbf{k}-\mathbf{k}_1).
\end{eqnarray}
Furthermore, using relation (\ref{mctnep}) between 
$f$ and $m$, we can re-write the above expression in such a way that it 
can be interpreted as a renormalized diagram. This diagram consists of
the left and right vertices given by $\tilde{v}_{\mathbf{k}}$, 
internal three-leg vertices
given by $\tilde{v}_{\mathbf{k}}/m(k)$, 
a four-leg vertex that represents an expression involving
$m_{\mathrm{MCT}}(k)$ and $m_{\mathrm{tMCT}}(k)$, 
and a bond equal to $f(k)S(k)$. We will comment on
the possible significance of this form of the above expression in Sec. \ref{discussion}.

Again, \textit{a priori}, it is not clear whether expression (\ref{bmct8b}) is positive or
negative. The explicit calculation described in the next subsection
suggests that expression (\ref{bmct8}) gives a significant, negative contribution 
to the irreducible memory function. 

\subsection{Perturbative calculation of the two 
corrections}\label{beyond3}

The two additional contributions to the irreducible memory function,
Eqs. (\ref{bmct4}) and (\ref{bmct7}), are functionals of the full response function. 
In principle, these contributions
can be added to the mode-coupling contribution, Eq. (\ref{mctexplicit}), 
and then the equation of motion for the response function,
\begin{eqnarray}\label{fulltime}
\lefteqn{ \int_0^t dt_1 \left[\delta(t-t_1)  \right. } 
\\ \nonumber && \left. +
M^{\mathrm{irr}}_{\mathrm{MCT}}(k;t-t_1) +
\delta M_1^{\mathrm{irr}}(k;t-t_1) + \delta M_2^{\mathrm{irr}}(k;t-t_1) \right]
\partial_{t_1} G(k;t_1) = -\frac{D_0 k^2}{S(k)} G(t)
\end{eqnarray}
can be
solved self-consistently. As the additional contributions
are expressed in terms of many-dimensional integrals (over wave-vectors and 
time) of the full response function, this procedure seems difficult and 
will not be attempted here. A somewhat easier task would be to consider the
self-consistent equation for the non-ergodicity parameter 
$f(k)=\lim_{t\to\infty} G(k;t)$, 
\begin{eqnarray}\label{fullnep}
\frac{f(k)}{1-f(k)} = m_{\mathrm{MCT}}(k) + \delta m_1(k) + \delta m_2(k),
\end{eqnarray}
where the functions at the right-hand-side are given by Eqs. (\ref{mmctdef}),
(\ref{bmct5a}) and (\ref{bmct8b}). 
This equation is a little more manageable because $\delta m_1$ and
$\delta m_2$ are functionals of the non-ergodicity parameter only. 
However, the full self-consistent solution of Eq. (\ref{fullnep}) still
seems rather difficult. 

To get some feeling regarding the size of the two additional terms, 
$\delta m_1$ and $\delta m_2$, contributing to 
the left-hand-side of Eq. (\ref{fullnep}) we will
calculate them perturbatively. Specifically, we will first solve
the standard mode-coupling equations for the non-ergodicity parameter, 
Eqs. (\ref{mctnep}) and (\ref{mmctdef}). Then, we will use the resulting 
mode-coupling non-ergodicity parameter to calculate the additional 
contributions. These contributions will be then compared to
the mode-coupling contribution $m_{\mathrm{MCT}}$. 

In order to perform numerical calculations we have to specify the system and its
state, and an approximate theory that will be used to calculate the 
static structure factor for this system. As in our earlier work \cite{GSPRL2003},
we will use the hard sphere system at the ergodicity-breaking 
transition point of the standard
mode-coupling theory and we will use the Verlet-Weiss approximation for the 
structure factor. We recall that using the Verlet-Weiss structure factor results in 
the ergodicity-breaking transition at volume fraction $\phi_{\mathrm{MCT}}=0.525$. 

\begin{figure}
\centerline{\includegraphics[scale=.4]{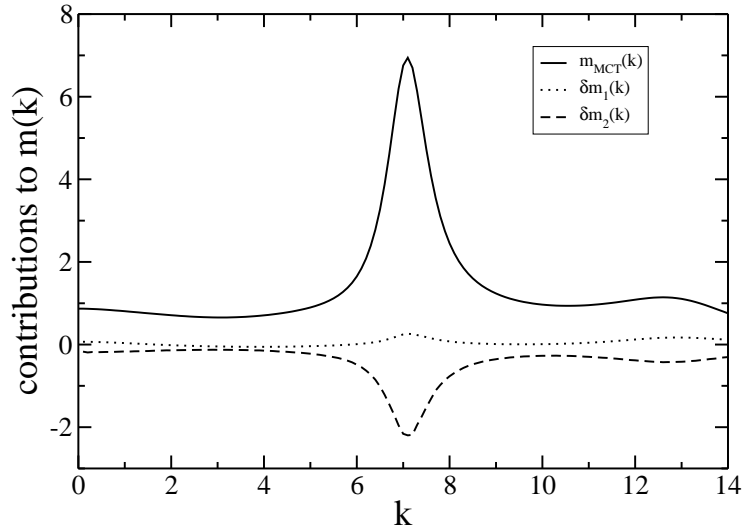}}
\caption{Contributions to the 
$m(k) = \lim_{t \to \infty} (S(k)/D_0 k^2) M^{\mathrm{irr}}(k;t)$, Eq. (\ref{mdef}),
calculated for a hard-sphere system using
Verlet-Weiss structure factor at the mode-coupling transition,
$\phi_{MCT} =0.525$. 
Solid line: mode-coupling contribution, $m_{\mathrm{MCT}}(k)$, 
Eq. (\ref{mmctdef}). 
Dotted line: the first correction, $\delta m_1(k)$, Eq. (\ref{bmct5a}).
Dashed line: the second correction $\delta m_2(k)$, Eq. (\ref{bmct8b}).
The corrections are calculated perturbatively, \textit{i.e.} using mode-coupling
theory's non-ergodicity parameter $f(k)$.}
\label{f:mresults}
\end{figure}

In Fig. \ref{f:mresults} we compare the mode-coupling result for function $m$,
$m_{\mathrm{MCT}}$ given by Eq. (\ref{mmctdef}), with two corrections, $\delta m_1$
given by Eq. (\ref{bmct5a}), and $\delta m_2$ given by Eq. (\ref{bmct8b}). 
We can see that the first correction is rather
small and, for most wave-vectors, positive. In contrast, the second correction
is more significant, with its magnitude reaching above 20\% of the mode-coupling
contribution, and negative. Thus, the second correction dominates and it likely
either moves the ergodicity breaking transition to higher volume fractions or 
removes it completely.

\section{Discussion}\label{discussion}

We have showed here that our earlier diagrammatic approach to the dynamics 
of fluctuations in equilibrium systems of interacting Brownian particles 
can be used to derive corrections to mode-coupling theory's irreducible 
memory function. We have presented explicit expressions for the two simplest corrections
and we have evaluated these corrections perturbatively. We found that one of these
corrections, which in our perturbative calculation gives a negative contribution 
to the irreducible memory function, is comparable to the mode-coupling contribution.
Thus, our results suggest that the simplest corrections are likely to 
move the ergodicity breaking transition to lower temperatures or higher volume
fractions. 

One important conclusion from our explicit calculations is that the easiest 
way to extend the standard mode-coupling theory is to concentrate on the 
self-consistent equation for the non-ergodicity parameter. This allows one
to avoid complications associated with the time dependence and reduces the 
technical complexity of the equations that need to be solved. 

Of course, while deriving approximate expressions for non-mode-coupling 
contributions to the irreducible memory function
one should strive to work with diagrams with bonds representing the
full response function. The second important conclusion from our calculations is 
that in order to avoid spurious, unphysical divergences one has to re-sum the
original diagrammatic expansion in such a way that restrictions on
intermediate time integrations are introduced.  

\begin{figure}
\centerline{\includegraphics[scale=.17]{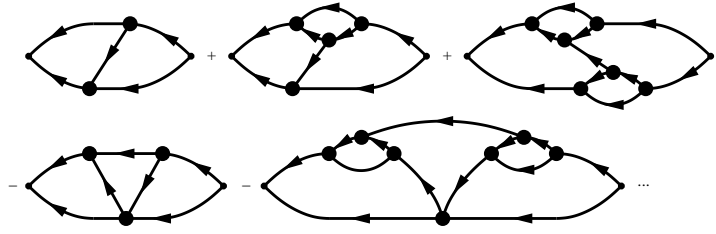}}
\caption{Re-summation of one particle irreducible, non-mode-coupling, 
fully renormalized diagrams 
with the following property: if one line is cut, the diagram either becomes
a mode-coupling diagram or a product of two mode-coupling diagrams.}
\label{f:disc1}
\end{figure}

\begin{figure}
\centerline{\includegraphics[scale=.17]{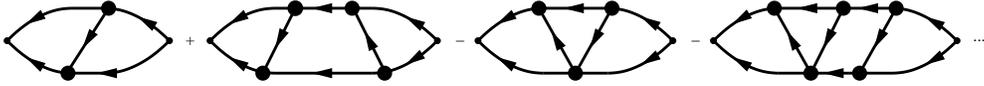}}
\caption{Re-summation of  one particle irreducible, non-mode-coupling, 
fully renormalized ladder diagrams where rungs of the ladders are non-mode-coupling
parts of the two contributions discussed in the present paper.}
\label{f:disc2}
\end{figure}

Our final expressions suggest that it should be possible to derive a fully 
renormalized diagrammatic series expansion for function $m(k)$ that is related to the 
long-time limit of the irreducible memory function,
Eq. (\ref{mdef}). In diagrams contributing to $m(k)$ the left and right vertices 
are given by modified vertex function $\tilde{v}_{\mathbf{k}}$, Eq. (\ref{vdef1}). 
The bonds represent the long-time limit of the full intermediate scattering
function, $\lim_{t\to\infty} F(k;t) \equiv f(k) S(k)$. The internal three-leg 
vertices represent modified vertex function $\tilde{v}_{\mathbf{k}}$ divided by $m(k)$.
Finally, the internal four-leg vertices represent a combination
of a product of two functions $\tilde{v}_{\mathbf{k}}$ divided by $m(k)$ and a novel 
term involving the transverse part of the memory matrix, $m_t(k)$. The internal
three-leg vertices pick up factors involving $m(k)$ and the internal four-leg vertices
pick up factors involving $m(k)$, and $m_t(k)$ as a result of 
re-summations that introduce restrictions on intermediate time integrations 
and thus remove spurious divergences. 
The fully renormalized diagrams which represent expressions (\ref{bmct5a}) and 
(\ref{bmct8b}) are the first and fourth diagrams in Fig. \ref{f:disc1} 
or the first and third diagrams in Fig. \ref{f:disc2}. 

In Fig. \ref{f:disc1} we show one possible extension of the work
presented here. In Secs. \ref{beyond1} and \ref{beyond2} we defined 
diagrams that were to be re-summed as, roughly speaking, mode-coupling
diagrams with one extra response function-like insertions and with 
additional conditions. Here we remove these additional conditions. 
As a result, in addition to the first and fourth diagrams in Fig. \ref{f:disc1},
we get a whole class of fully renormalized diagrams, some of which are 
showed in Fig. \ref{f:disc1}. Preliminary results \cite{GSHHEF} suggest that
if these diagrams are re-summed perturbatively (\textit{i.e.} if 
mode-coupling $f$, $m$ and $m_t$ are used instead of the exact functions),
the sum of these diagrams diverges upon approaching the ergodicity
breaking transition of the standard mode-coupling theory. The strength of 
the divergence depends on the dimensionality of the system and the divergence  
vanishes in high enough dimension. The analysis of this divergence should allow
us to calculate the upper critical dimension of the mode-coupling theory.
An analogous calculation in the framework of the static replica approach
appeared recently \cite{FJPUZ}. 

In Fig. \ref{f:disc2} we show the second possible extension. Roughly speaking,
we propose to re-sum a series of fully renormalized ladder diagrams where rungs 
of the ladders are represented by non-mode-coupling parts of the two contributions 
discussed in this paper (which are represented by the first and third diagrams in 
Fig. \ref{f:disc2}). This re-summation could be combined with Eq. (\ref{mctnep}) 
resulting in a self-consistent calculation. 

Finally, we could also attempt to use the fully renormalized diagrammatic
series to derive a self-consistent equation for a vertex function. 

\section*{Acknowledgments}
Part of this work was done when I was a Visiting Professor at
Yukawa Institute for Theoretical Physics in Kyoto. I would like to acknowledge
the Institute for its generous hospitality. I am grateful to Hisao Hayakawa 
for many insightful discussions during my stay in Kyoto and for the invitation 
to write this paper. I thank Elijah Flenner, Matthias Fuchs, Hisao Hayakawa 
and Koshiro Suzuki for their comments on the manuscript. 
Support by NSF Grant CHE 0909676 is gratefully acknowledged.

\end{document}